\documentclass[aps,prd,twocolumn,superscriptaddress,amssymb]{revtex4-2}
\usepackage{bm}
\usepackage{hyperref}%
\usepackage{amsmath}
\usepackage{siunitx}
\usepackage{graphicx}	
\usepackage{aas_macros}
\usepackage{xcolor}
\usepackage[normalem]{ulem}
\usepackage[cal=boondox]{mathalfa}

\begin{document}
	
	
	\title{Neutrino-electron magnetohydrodynamics in an expanding Universe}
	
	
	\author{L. M. Perrone}
	\email[]{lmp61@cam.ac.uk}
	\affiliation{Department of Applied Mathematics and Theoretical Physics, University of Cambridge, Wilberforce Road, Cambridge CB3 0WA, UK}
	\affiliation{Department of Physics, University of Oxford, Parks Road, Oxford OX1 3PU, UK}
	
	\author{G. Gregori}
	\affiliation{Department of Physics, University of Oxford, Parks Road, Oxford OX1 3PU, UK}
	
	\author{B. Reville}
	\affiliation{Max-Planck-Institut f\"{u}r Kernphysik, Saupfercheckweg 1, D-69117 Heidelberg, Germany}
	
	\author{L. O. Silva}
	\affiliation{GoLP/Instituto de Plasmas e Fus\~{a}o Nuclear, Instituto Superior T\'{e}cnico, Universidade de Lisboa, 1049-001 Lisboa, Portugal}
	
	\author{R. Bingham}
	\affiliation{Rutherford Appleton Laboratory, Chilton, Didcot, Oxon OX11 OQX, UK}
	\affiliation{Department of Physics, University of Strathclyde, Glasgow G4 0NG, UK}
	
	
	\date{\today}
	
	\begin{abstract}
		We derive a new model for neutrino-plasma interactions in an expanding universe that incorporates the collective effects of the neutrinos on the plasma constituents. We start from the kinetic description of a multi-species plasma in the flat Friedmann-Robertson-Walker metric, where the particles are coupled to neutrinos through the charged- and neutral-current forms of the weak interaction. We then derive the fluid equations and specialize our model to (a) the lepton epoch, where we consider a pair electron-positron plasma interacting with electron (anti-)neutrinos, and (b) after the electron-positron annihilation, where we model an electron-proton plasma and take the limit of slow ions and inertia-less electrons to obtain a set of neutrino-electron magnetohydrodynamics (NEMHD) equations. In both models, the dynamics of the plasma is affected by the neutrino motion through a ponderomotive force and, as a result, new terms appear in the induction equation that can act as a source for magnetic field generation in the early universe. A brief discussion on the possible applications of our model is proposed.
	\end{abstract}
	
	
	\maketitle
	
	\section{Introduction}

	Neutrinos play an important role in many astrophysical contexts \cite{Bahcall1989}: they are produced in abundance in core-collapse supernovae \cite{Bionta1987,Hirata1987}, and by thermonuclear reactions in the interiors of stars \cite{Bahcall1972}; moreover, relic neutrinos that originated in the early universe are thought to permeate all space and can provide valuable information on the cosmology of the Big Bang \cite{Steigman2012}. Neutrinos interact with other plasma constituents through the electroweak force and its associated charged and neutral currents. Even though the interaction cross-section of the neutrinos with matter is extremely small, the many-body interaction between neutrinos and electrons can drive plasma instabilities \cite{Bingham1994,Bingham1996,Silva1999E,Silva1999D,Silva1999L,Silva2000,Bingham2004} and generate strong magnetic fields \cite{Shukla1997,Shukla1998,Brizard2000}. The self-consistent generation of electromagnetic fields through collective interactions  is of particular relevance to the problem of cosmic magnetogenesis \cite{Turner1988,Kronberg1994,Enqvist1998,Son1999,Giovannini2004}, and may offer a possible solution to the generation of primordial magnetic fields in the early universe.\\

	Numerous works have studied previously the collective interactions between neutrinos and plasma particles, adopting both quantum  and semi-classical neutrino descriptions. Quantum approaches have focused on the effect of processes such as the lepton neutrino-antineutrino asymmetry \cite{Dolgov2002,Semikoz2004,Dvornikov2014} as well as axial-vector coupling \cite{Yamamoto2016,Pandey2020} on the induction of electric currents in the plasma. Approaches in the semi-classical limit have modelled the neutrino population both in a kinetic framework \cite{Semikoz1987,Silva2000,Silva2000b} or as an ideal fluid \cite{Brizard1999,Shukla2003,Haas2016}. In the semi-classical approximation, the effects of neutrino-electron nonlinear interactions can be included in magnetohydrodynamic (MHD) models by making use of the formal analogy between the electromagnetic and weak interaction \cite{Silva2000b}. The resulting equations differ from standard MHD in that a new \textit{ponderomotive} force, representing the collective force of the neutrinos on plasma particles, appears in the momentum and induction equations \cite{Brizard2000,Haas2016}. This additional term depends on the number density and the velocity of neutrinos, but not on the magnetic field strength, and can therefore generate a primordial magnetic seed.\\

	An appropriate framework for the study of cosmic magnetogenesis through the collective interaction of neutrinos with the plasma, however, should take into account the expansion of the universe, captured in the scale factor $a(t)$. Thus, the main purpose of this paper is to develop simplified fluid models of neutrino-plasma interactions in comoving coordinates, that can be employed for further analytical or numerical studies. The plan of the paper is as follows: in Section~\ref{sec:theory_prelim}, we review the physical model of neutrino-plasma interactions in the semiclassical limit, recasting it in a form suitable to our calculations, and introduce our choice of metric; in Section~\ref{sec:physical_model} we derive the equations of a plasma made of charged particles and neutrinos in an expanding universe coupled to Maxwell's equations (\ref{sec:maxwell}). Using conformal coordinates, we start with a kinetic framework (\ref{sec:kinetic_eqs}), then take the hydrodynamic limit (\ref{sec:fluid_equations}). Finally, we develop a reduced model of an electron-positron plasma (Section~\ref{sec:NEP-MHD}) and an electron-proton plasma (Section~\ref{sec:NE-MHD}) interacting with neutrinos, and discuss potential applications (Section~\ref{sec:discussion}).
	
	\section{Theoretical preliminaries}\label{sec:theory_prelim}
	
	\subsection{Neutrino-Plasma interaction}\label{sec:electron_neutrino_int}
	
	In this work we model the plasma-neutrino interactions based on a semi-classical treatment of the interaction Lagrangian between neutrinos and the background particles \cite{Wolfenstein1978,Tajima2002}. The semiclassical approximation discards the axial-vector contribution in the interaction Lagrangian, which is associated with the spin of the fermions, and requires the de Broglie wavelength of the neutrino ($\lambda_{\nu}$) to be much shorter than the plasma skin depth $d_e = c/ \omega_{pe}$, with $\omega_{pe} = (4 \pi n_{e} e^2 / \gamma_{e} m_{e} )^{1/2}$ the electron plasma frequency, which is the typical length scale of perturbations in the plasma, $n_{e}$ is the rest number density of the hot electrons, and $\gamma_e m_e$ is their relativistic mass. We will adopt the approximation of the neutrino forward
	scattering on plasma particles, focusing on the collective, i.e. \textit{mean-field}, effects of the neutrino distribution.\\

	The interaction Lagrangian\footnote{Note that we use the definition of $\mathcal{L}$ as the time derivative of the action $S$ with respect to coordinate time and not proper time.} for a single neutrino of type $\nu$ with velocity $\bm{v}_{\nu}$ in a background made of particles of species $s$ (where $s$ refers to either electrons, protons or neutrons) is \cite{Tajima2002,Silva2000}
	\begin{eqnarray}\label{eq:int_lagr}
		\mathcal{L}_{int,\nu s}^{(W)} = -q_{s\nu}^{(W)} \left(n_s - \frac{\bm{N}_s \cdot \bm{v}_{\nu}}{c^2}\right) ,
	\end{eqnarray}
	where $q_{\nu s}^{(W)}$ is the effective charge of the weak interaction, and $n_s$ and $ \bm{N}_s$ are the number and current density of species $s$ in the laboratory frame, respectively. The effective charge is proportional to the Fermi constant of weak interaction $G_F$ and satisfies the following \cite{Brizard2000}:
	\begin{align}\label{eq:charge_identities}
		q_{s\nu}^{(W)} = - q_{\bar{s}\nu}^{(W)} = - q_{s\bar{\nu}}^{(W)} = q_{\bar{s}\bar{\nu}}^{(W)},
	\end{align}
	where the bar denotes the corresponding antiparticle. In particular, for electrons and nucleons the effective charge takes the following values \cite{Brizard1999}
	\begin{align}
		q_{s\nu}^{(W)} = \begin{cases}
			\sqrt{2} G_F \hfill  \, \, \, \, \, s=e,\\
			0 \hfill  \, \, \, \, \, s=p, \\
			-G_F / \sqrt{2} \hfill  \, \, \, \, \, s=n.
		\end{cases} 
	\end{align} 
	Finally, if heavier ions (made of $P$ protons, $N$ neutrons and $P-Z$ electrons) are present in the plasma, the effective charge becomes instead $G_F \left[2(P-Z)-N\right] /\sqrt{2}$ \cite{Brizard1999}.
	
	The semiclassical neutrino-particle interaction Lagrangian in Eq.~\eqref{eq:int_lagr} is formally identical to that of a charged particle in the presence of an electro-magnetic field \cite{Silva2000} if we define the 
	generalized 4-potential $A_s^{\alpha} =  (n_s,c^{-1} \bm{N}_s)$ of the "weak" interaction, so that for a flat spacetime $\mathcal{L}_{int,s\nu}^{(W)}$ can be written as
	\begin{eqnarray}\label{eq:int_lagr_weak_cov}
		\mathcal{L}_{int,s\nu}^{(W)} = -\frac{q^W_{s\nu}}{c} \eta_{\alpha \beta} A_s^{\alpha} \frac{d x^{\beta}_{\nu}}{d t},
	\end{eqnarray}
	with $\eta_{\alpha \beta} = \text{diag}(1,-1,-1,-1)$ the Minkowski metric, $x^{\alpha}_{\nu}$ the 4-position of the neutrino, and where we used Einstein's convention for summation over repeated indices.\\
	
	In addition to the force induced by the background electrons on the neutrinos, the neutrino distribution can itself affect the motion of the electrons through a macroscopic ponderomotive force \cite{Silva1999E,Silva1999D,Silva1999L}. This effect is analogous to the force exerted by an electromagnetic pulse on the plasma, as is the case in laser wakefield acceleration \cite{Tajima1979,Esarey2009} (see \cite{Gibbon2005,Mulser2010} for a general reference). Interestingly, the interaction of a single particle of species $s$ with a background of neutrinos can be expressed in terms of a Lagrangian which shares the same formal structure as Eq.~\eqref{eq:int_lagr_weak_cov}, where the effective 4-potential $A_{\nu}^{\alpha}$ is now proportional to the neutrino number current $A_{\nu}^{\alpha} =  (n_{\nu},c^{-1} \bm{N}_{\nu})$, yielding
	\begin{eqnarray}\label{eq:int_lagr_pondero_cov}
		\mathcal{L}_{int,s\nu}^{(P)} = -\frac{q^W_{s\nu}}{c} \eta_{\alpha \beta} A_{\nu}^{\alpha} \frac{d x^{\beta}_{s}}{d t},
	\end{eqnarray}
	with $x^{\alpha}_{s}$ now the 4-position of particle $s$, and where $n_{\nu}, \bm N_{\nu}$ are the number and current density of the neutrinos in the laboratory frame.\\
	
	We can further exploit the analogy with electromagnetic interaction, and define the weak field tensor $F_s^{\mu \nu}$ in a similar fashion to the usual EM field tensor, i.e. $F_{s}^{\alpha \beta} = \partial^{\alpha} A_{s}^{\beta} - \partial^{\beta} A_s^{\alpha}$. In the same spirit, we introduce the effective electric and magnetic fields for the weak force field interaction, as
	\begin{eqnarray}
		F^{i0}_s = -\nabla n_s - \frac{1}{c^2} \frac{\partial \bm N_s}{\partial t} = \bm E_s, \label{eq:effective_weak_electric}\\
		F^{ij}_s = - \epsilon^{kij} \left(\nabla \times \bm N_s/c  \right)_{k}= - \epsilon^{kij} (\bm B_s)_k , \label{eq:effective_weak_magnetic}
	\end{eqnarray}
	and analogously for the ponderomotive potential $F_{\nu}^{\alpha \beta} = \partial^{\alpha} A_{\nu}^{\beta} - \partial^{\beta} A_{\nu}^{\alpha}$, with the definitions of the effective fields $\bm E_{\nu}, \bm B_{\nu}$ as in Eq.~\eqref{eq:effective_weak_electric}-\eqref{eq:effective_weak_magnetic} with the neutrino current replacing the electron current. As we shall see in Section \ref{sec:kinetic_eqs}, the introduction of the effective electric and magnetic fields for the weak and ponderomotive force allows us to extend the general relativistic Vlasov-Maxwell system of equations to take into account neutrino interactions in a very straightforward manner.

	\subsubsection{Validity of our model}

\begin{figure}
		\centering
		\includegraphics[width=1.0\columnwidth]{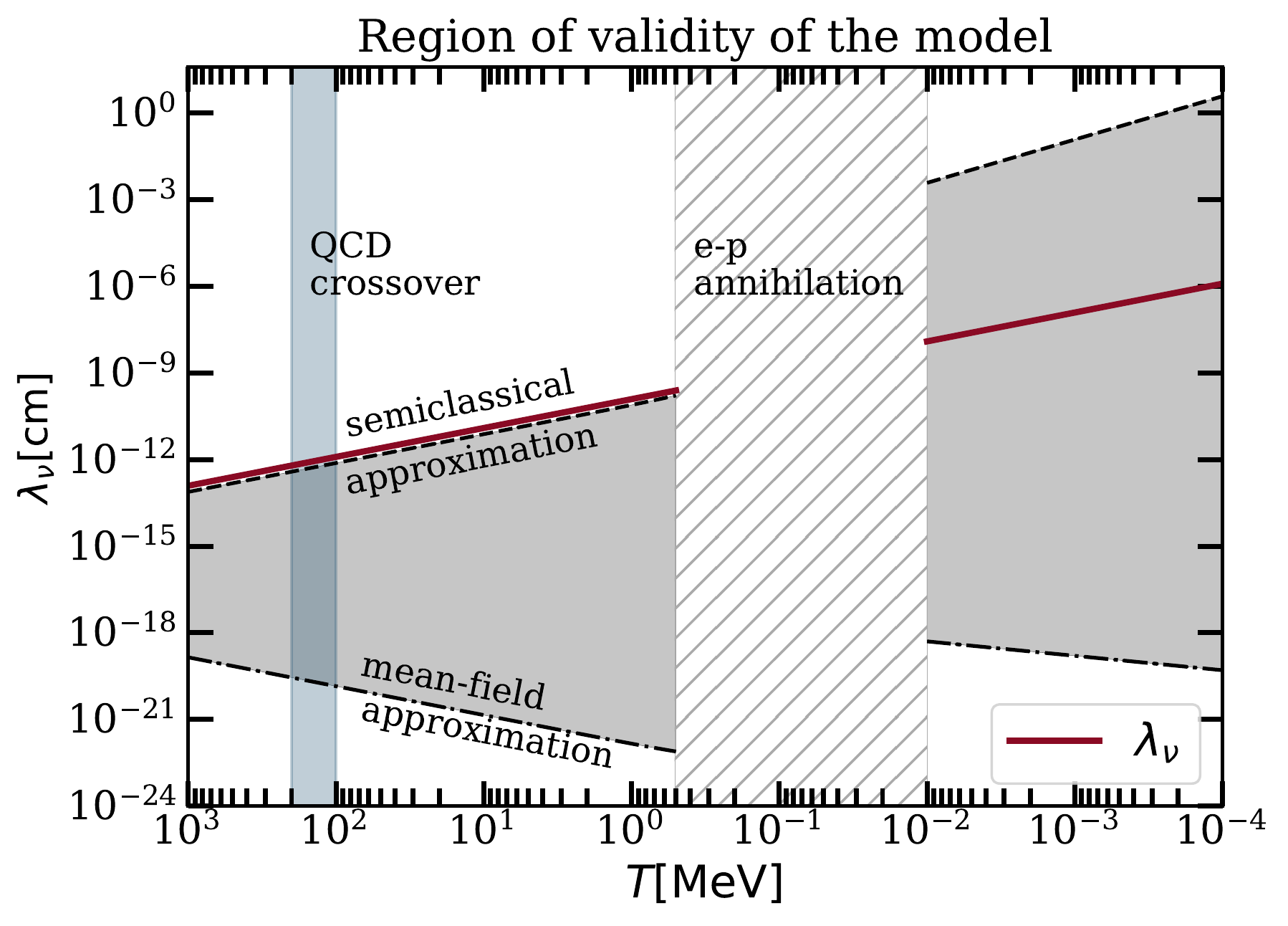}
		\caption{De Broglie wavelength ($\lambda_{\nu}$) of the neutrinos in the early universe (solid red line). The shaded area represents the region of validity of the semiclassical (black dashed line) and mean-field approximations (black dash-dotted line), while the hatched area denotes the electron-positron annihilation. Finally, the region shaded in blue indicates the time of the QCD crossover. Our model is marginally applicable for $T \gtrsim \si{MeV}$, while for $T \lesssim 10 \,\ \si{keV}$ the semiclassical and mean-field approximations are well respected.}
		\label{fig:semiclassical_approximation}
	\end{figure}

	If we look at the early Universe, we find that the semi-classical approximation is marginally verified for $T \gtrsim \si{MeV}$, where the ratio between the neutrino de Broglie wavelength and the plasma skin-depth takes the simple expression $\lambda_{\nu} / d_e \simeq 0.95 g^*_e(T)^{1/2}$, with $g^*_e(T)$ the effective number of degrees of freedom of the electrons. For $T \gtrsim \si{MeV}$, $g^*_e(T)$ does not vary significantly with temperature, and therefore we have $\lambda / d_e \simeq 1.5-1.6 $ for a wide range of temperatures from the QCD crossover ($T\sim 100 \,\, \si{MeV}$) to the end of the lepton epoch ($T\sim 0.5 \,\, \si{MeV}$). After the electron-positron annihilation, only a small number of electrons are left, and the electron skin depth becomes much larger than the neutrino de Broglie wavelength, see Fig.~\ref{fig:semiclassical_approximation}. Through a similar argument, it can be shown that for $T \gtrsim \si{MeV}$ the de Broglie wavelength of the electrons/positrons is of the same order of the inter-particle distance $d_n = n_e^{-1/3}$. While in this regime quantum effects start to become important, we treat the pair plasma as non-degenerate for simplicity.\\

	To formalize the validity of the mean-field approximation we also compare the force on the electrons due to collisions with neutrinos $\bm F_{coll} \sim \sigma_{\nu e} n_{\nu} \bm p_{\nu} c$, where $ \sigma_{\nu e} $ is the scattering cross-section and $\bm p_{\nu}$ the relativistic momentum of the neutrinos, with the ponderomotive force due to collective neutrino-electron interactions $\bm F_{\nu} \sim \sqrt{2}G_F \nabla n_{\nu}$. Assuming that the scale of variation of the neutrino number density is of the order of the plasma skin depth, the condition $| \bm F_{coll}| \ll | \bm F_{\nu} |$ requires that
	\begin{align}
		\frac{\lambda_{\nu}}{d_n} \gg 0.032 G_F^0 T^2 \gamma_e ^{1/2} \left( \frac{d_n}{r_e} \right)^{1/2}, \label{eq:semiclassical_cond1}
	\end{align}
	where $G_F^0 = G_F/(\hbar c)^3$, $\gamma_e$ is the relativistic Lorentz factor for the electrons, and $r_e$ is the classical radius of the electron. In other words, for the collective effects due to the ponderomotive force to prevail over particle scattering, the de Broglie wavelength of the electron has to be greater than a certain minimum value which depends on temperature and is proportional to the inter-particle distance (note, however, that the right-hand side of Eq.~\eqref{eq:semiclassical_cond1} is extremely small, with values in the range of $10^{-12}-10^{-6}$ for the temperatures of interest). At the QCD transition, when the electrons and the neutrinos are in thermal equilibrium with the radiation field, this condition is largely satisfied, and the collisional force is $10^{-8}$ weaker than the ponderomotive force. After the electron-positron annihilation, the magnitude of the collisional forces further decreases (see Fig.~\ref{fig:semiclassical_approximation}). 

	\subsection{The FRW metric}
	
	The metric of the spacetime considered in this work is the spatially flat Friedmann-Robertson-Walker (FRW) metric
	\begin{eqnarray}
		d s^2 =  dt^2 - a^2 (t) \sum_{i=1,2,3} (d x^i)^2,
	\end{eqnarray}
	with $t$ the cosmological time, $x_i$ the comoving spatial coordinates, and $a(t)$ the scale factor normalized so that $a=1$ at present time. The metric can be further simplified by introducing a conformal time coordinate $\eta$ defined as $d \eta = a^{-1}(t) d t$, so that $g_{\mu \nu}$ can be recast in the form
	\begin{eqnarray}\label{eq:frw_metric}
		g_{\mu \nu} = a^2 (\eta) \times \text{diag} \left(1,-1,-1,-1\right).
	\end{eqnarray}
	
	The only non-zero Christoffel symbols of the flat FRW metric in Eq.~(\ref{eq:frw_metric}) are
	\begin{eqnarray}
		\Gamma^i_{0i} = \Gamma^i_{i0} = \Gamma^0_{\mu \mu} = \frac{a'}{a} = \dot{a},  
	\end{eqnarray}
	where the indices can take values of $i=1,2,3$ and $\mu=0,1,2,3$, and where we used a prime (dot) to indicate differentiation with respect to conformal (cosmological) time.

	\section{Derivation of the model}\label{sec:physical_model}
	
	Our aim is to develop a simplified model for neutrino-plasma interactions in an expanding space-time described by the flat FRW metric, which we assume to be externally fixed. In order to do so, we proceed by first deriving a system of kinetic equations for the ions, electrons and neutrinos, that are coupled to Maxwell's equations. We then integrate the kinetic equations in momentum space to obtain a set of fluid equations in comoving coordinates. \\
	
	To avoid confusion, in the next sections we will be careful to distinguish quantities defined with respect to the FRW metric in Eq.~\eqref{eq:frw_metric}, from those defined in a flat space-time (the "laboratory" frame), which will be denoted by carets. Furthermore, we use natural units with $c=1$.\\
	
	\subsection{Maxwell's equations}\label{sec:maxwell}
	
	Maxwell's equations in the FRW metric have been derived in numerous works before  -- both in "3+1" split formalism, see e.g. \cite{Thorne1982,Holcomb1989}, and in covariant formulation --  we will therefore only give a brief summary below. In covariant form, Maxwell's equations read
	\begin{eqnarray}\label{eq:maxwell_covariant}
		F^{\mu \nu}_{;\mu} = 4\pi J^{\nu}, \quad F_{\left[ \alpha \beta ; \gamma \right]} = 0,
	\end{eqnarray}
	where $F^{\mu \nu}$ and $J^{\mu}$ are the electro-magnetic field tensor and the electric 4-current, respectively, the semicolon operator represents the covariant derivative and where we have used the antisymmetric tensor notation to express the Gauss-Faraday law. For the flat FRW metric in Eq.~\eqref{eq:frw_metric}, we obtain
	\begin{eqnarray}
		\frac{\partial}{\partial \eta} \left( a^2 \hat{\bm E} \right) = \nabla \times \left( a^2 \hat{\bm B} \right) - 4 \pi \left( a^3 \hat{\bm J} \right) , \label{eq:max_frw_1} \\
		\frac{\partial}{\partial \eta} \left( a^2 \hat{\bm B} \right) = - \nabla \times \left( a^2 \hat{\bm E} \right) , \label{eq:max_frw_2} \\
		\nabla \cdot \left( a^2 \hat{\bm E} \right) = 4 \pi \left( a^3 \hat{\rho} \right), \quad 	 \nabla \cdot \left( a^2 \hat{\bm B }\right) = 0, \label{eq:max_frw_3}
	\end{eqnarray}
	where the spatial derivatives are taken with respect to comoving coordinates, $\hat{\bm{E}},\hat{\bm{B}}$ are the flat-space electric and magnetic fields, and $(\hat{\rho},\hat{\bm J})$ the flat-space electric current density. As is well known, Eqs.~\eqref{eq:max_frw_1}-\eqref{eq:max_frw_3} are formally identical to the standard electrodynamics equations in Minkowski space (to which they reduce to for $a(t)=1$) if we define the following \textit{conformal} quantities
	\begin{eqnarray}\label{eq:conformal_fields}
		\mathbcal{E} = a^2 \hat{\bm E}, \quad \mathbcal{B} = a^2 \hat{\bm B}, \quad \mathbcal{J} = a^3 \hat{\bm J}, \quad \varrho = a^3 \hat{\rho}.
	\end{eqnarray}

	\subsection{Collisionless kinetic equations}\label{sec:kinetic_eqs}
	
	In this section, we derive the kinetic equations for the ions, electrons and neutrinos in the flat FRW metric. We note that a different derivation of the relativistic neutrino kinetic equations using methods from Finite Temperature Quantum Field Theory can be found in Ref. \cite{Semikoz1987} for the Minkowski space-time. For a generic particle species $s$, the evolution of the single-particle distribution function $f_s$ in the absence of collisions is given by the general relativistic Liouville's equation \cite{Lifshitz:99987,Ehlers1971}
	\begin{eqnarray}\label{eq:GR_RKE}
		\frac{d f_s}{d \tau} = g_{\alpha \beta}\frac{d x^{\alpha}}{d \tau} \frac{\partial f_s}{\partial x_{\beta}} + g_{\alpha \beta} \frac{d p^{\alpha}}{d \tau} \frac{\partial f_s}{\partial p_{\beta}} = 0,
	\end{eqnarray}
	where $\tau$ is the proper time and $p^{\alpha}$ the 4-momentum. In Eq.~\eqref{eq:GR_RKE} $d x^{\alpha} / d \tau$ and $d p^{\alpha} / d \tau$ are determined through the equations of motion of the particle, and ensure that we stay on the hypersurface defined $m=const$ of the 8-dimensional phase space  $(x^{\alpha},p^{\alpha})$ \cite{Ehlers1971}. For instance, the trajectory of a particle with mass $m$ and electric charge $q$ that moves in a fixed gravitational field in presence of an electromagnetic field is given by the usual Lorentz-Einstein equations \cite{Ehlers1971}
	\begin{eqnarray}\label{eq:lorentz_einstein}
		\frac{d x^{\mu}}{d \tau} = \frac{p^{\mu}}{m},  \quad \frac{d p^{\mu}}{d \tau} + \Gamma^{\mu}_{\lambda \nu} p^{\lambda} u^{\nu} = q g_{\alpha \nu} F^{\mu \nu}  u^{\alpha}, 
	\end{eqnarray}
	where $u^{\alpha} = p^{\alpha}/m$. Using the formal analogy between the EM force and the weak force explored in Section \ref{sec:electron_neutrino_int}, it is straightforward to extend Newton's law to account for the neutrino interaction in the equations of motion  by simply adding the term $\sum_{\nu} q^W_{s\nu} g_{\alpha \beta} F^{\mu \beta}_{\nu}  u^{\alpha}_s$ (where the sum is over the neutrino/antineutrino flavours and their corresponding antiparticle) on the right-hand side of Eq.~(\ref{eq:lorentz_einstein}) for the plasma constituents, and replacing the Lorentz force with $\sum_s q^W_{s\nu} g_{\alpha \beta} F^{\mu \beta}_{s}  u^{\alpha}_{\nu}$ in the case of neutrinos, where the sum now runs over the particle species and their corresponding antiparticle.

	In the flat FRW metric, Eqs.~\eqref{eq:GR_RKE}-\eqref{eq:lorentz_einstein} can be considerably simplified if we introduce the conformal momentum $\mathcal{p}^{\alpha} = a^2 p^{\alpha}$($=a \hat{p}^{\alpha}$), which allows us to rewrite the kinetic equation as
	\begin{align}\label{eq:vlasov_frw}
		\left[ \frac{\partial }{\partial \eta} +  \hat{ \bm v} \cdot \nabla + \frac{d \mathbcal{p}}{d \eta} \cdot \frac{\partial}{\partial \mathbcal{p}} \right] f_s (\bm x, \mathbcal{p}, \eta) = 0,
	\end{align}
	where the gradient is in comoving coordinates $\bm x$, and where we used the chain rule to express the derivative with respect to the space component of the conformal momentum $\mathbcal{p}$ in the kinetic equation. In terms of the new coordinates, the force acting on the ions/electrons and neutrinos becomes, respectively:
	\begin{align}
		\frac{d \mathbcal{p}}{d \eta} = q_s \left(\mathbcal{E} +  \hat{ \bm v}_s \times \mathbcal{B} \right) + \sum_{\nu} q^W_{s\nu} \left(\mathbcal{E}_{\nu} +  \hat{ \bm v}_s \times \mathbcal{B}_{\nu} \right), \label{eq:lorentz_frw_electrons_ions}\\
		\frac{d \mathbcal{p}}{d \eta} = \sum_s q^W_{s\nu} \left(\mathbcal{E}_s +  \hat{ \bm v}_{\nu} \times \mathbcal{B}_s \right), \label{eq:lorentz_frw_neutrinos}
	\end{align}
	with $q_i = Z_i e$ ($q_e = -e$) for the ions (electrons), and where the conformal effective fields of the weak interaction and of the ponderomotive force have been introduced analogously to Eq.~\eqref{eq:conformal_fields}, i.e. $	\mathbcal{E}_{s} = a^2 \hat{\bm E}_{s} $, $\mathbcal{B}_s = a^2 \hat{\bm B}_{s}$ and so on for  $\mathbcal{E}_{\nu}, \mathbcal{B}_{\nu}$. We note that in Eq.~\eqref{eq:vlasov_frw}-\eqref{eq:lorentz_frw_neutrinos}, the flat-space velocity variable $\hat{\bm v}$ is related to the conformal momentum  $\mathbcal{p}$ by the relation
	\begin{align}
		\hat{\bm v} = \frac{\mathbcal{p}}{\mathcal{p}^0} = \frac{\mathbcal{p}}{\sqrt{\mathbcal{p}^2 + m^2 a^2(\eta)}}.
	\end{align}
	Without the neutrino interaction terms, Eq.~\eqref{eq:vlasov_frw}-\eqref{eq:lorentz_frw_neutrinos} reduce to the equation of motion of a charged particle in conformal coordinates \cite{Dettmann1993}.\\
	
	Together with Maxwell's equations derived in Section \ref{sec:maxwell}, Eq.~\eqref{eq:vlasov_frw}-\eqref{eq:lorentz_frw_neutrinos} constitute the relativistic Vlasov-Maxwell equations for neutrino-plasma interactions, with the system closed by the definition of the conformal charge and current density in terms of the single-particle distribution function
	\begin{align}
		\varrho = \sum_{s=i,e} q_s \int d^3 \mathcal{p}  f_s , \quad \mathbcal{J} = \sum_{s=i,e} q_s \int \frac{d^3 \mathcal{p}}{\mathcal{p}^0} \mathbcal{p} f_s .
	\end{align}
	
	\subsection{Fluid equations}\label{sec:fluid_equations}
	
	To derive the fluid equations of the plasma and the neutrino populations, we integrate Eq.~\eqref{eq:vlasov_frw} in conformal momentum space, treating $(\eta, \bm x, \mathbcal{p})$ as independent variables. We therefore define the conformal number density and stress-energy tensor for the species $s$ as
	\begin{align}
		\mathcal{N}_s^{\alpha} = \int \frac{d^3 \mathcal{p}}{\mathcal{p}^0} \mathcal{p}^{\alpha} f_s, \quad \mathcal{T}_s^{\alpha \beta} = \int \frac{d^3 \mathcal{p}}{\mathcal{p}^0} \mathcal{p}^{\alpha} \mathcal{p}^{\beta} f_s ,
	\end{align}
	that are related to the corresponding flat-space quantities as $\mathcal{N}_s^{\alpha}= a^3  \hat{N}_s^{\alpha}$ and $\mathcal{T}_s^{\alpha \beta}= a^4 \hat{T}_s^{\mu \nu}$. The equations for number, momentum and energy conservation for each particle species can thus be derived  multiplying Eq.~\eqref{eq:vlasov_frw} by $1$, $\mathbcal{p}$ and $\mathcal{p}^0$, respectively, and integrating in $d^3 \mathcal{p}$. Note that $\mathcal{p}^0$, is not an independent variable, but is instead defined in terms of $(\mathbcal{p},\eta)$. Indeed, it is easily shown that
	\begin{align}
		\frac{\partial \mathcal{p}^0}{\partial \eta} = \frac{a' \mathcal{p}^0}{a}   \left( 1-\frac{ \mathbcal{p}^2}{ (\mathcal{p}^0)^2} \right), \quad \frac{\partial \mathcal{p}^0}{\partial \mathbcal{p}} = \frac{\mathbcal{p}}{\mathcal{p}^0}.
	\end{align}
	
	We now assume that the ions, electrons and neutrinos can be represented as perfect fluids, ignoring effects due to viscosity and heat conduction. For a perfect fluid, the conformal stress-energy tensor takes the simple form \cite{DeGroot1980}
	\begin{align}
		\mathcal{T}^{\alpha \beta} = (\mathcal{e} + \mathcal{P}) \hat{U}^{\alpha} \hat{U}^{\beta} - \mathcal{P}  \eta^{\alpha \beta},
	\end{align}
	where $\hat{U}^{\alpha}=(\gamma, \gamma \hat{\bm u})$ is the hydrodynamic 4-velocity of the fluid for a flat-space observer (we follow Eckart's convention for the definition of $\hat{U}^{\alpha}$), $\eta^{\alpha \beta}$ is the Minkowski metric, and where we introduced the total (rest plus internal) conformal energy density $\mathcal{e}$ and the conformal pressure $\mathcal{P}$ in the rest-frame of the fluid, that differ from their ordinary flat-space counterparts by a factor of $a^4$. Similarly, the conformal current density is expressed in terms of the hydrodynamic 4-velocity as $\mathcal{N}^{\alpha} = \mathcal{n} \hat{U}^{\alpha}$, where $\mathcal{n}$ is the conformal number density in the rest frame of the fluid.
	
	With these definitions, the continuity, momentum and energy equations for the plasma species read
	\begin{align}
		&\frac{\partial (\gamma_s \mathcal{n}_s )}{\partial \eta} + \nabla \cdot \left(\gamma_s \mathcal{n}_s \hat{\bm u}_s\right) = 0, \label{eq:fluid_charged_1} \\
		&\frac{\partial \left[ \gamma_s^2 (\mathcal{e}_s + \mathcal{P}_s) \hat{\bm u}_s \right]}{\partial \eta} + \nabla \cdot \left[ \gamma_s^2 (\mathcal{e}_s + \mathcal{P}_s) \hat{\bm u}_s \hat{ \bm u}_s + \mathcal{P}_s \openone \right] \label{eq:fluid_charged_2} \\ 
		&= q_s \gamma_s \mathcal{n}_s \left( \mathbcal{E} + \hat{ \bm u}_s \times \mathbcal{B} \right) + \sum_{\nu} q^W_{s\nu} \gamma_s \mathcal{n}_s \left( \mathbcal{E}_{\nu} + \hat{ \bm u}_s \times \mathbcal{B}_{\nu} \right) ,  \nonumber \\
		&\frac{ \partial \left[\gamma_s^2 (\mathcal{e}_s + \mathcal{P}_s \hat{u}_s^2) \right]}{\partial \eta}  + \nabla \cdot \left[ \gamma_s^2 (\mathcal{e}_s + \mathcal{P}_s) \hat{\bm u}_s \right] \label{eq:fluid_charged_3} \\ 
		&=  - \frac{a'}{a}  (3\mathcal{P}_s - \mathcal{e}_s)  + q_s \gamma_s \mathcal{n}_s \hat{ \bm u}_s \cdot \mathbcal{E} + \sum_{\nu} q^W_{s\nu} \gamma_s \mathcal{n}_s  \hat{ \bm u}_s \cdot \mathbcal{E}_{\nu}  , \nonumber
	\end{align}
	which reduce to those derived in \cite{Dettmann1993} if we discard the neutrino interaction terms. The neutrino fluid instead obeys the following momentum and energy equations
	\begin{align}
		&\frac{\partial \left[ \gamma_{\nu}^2 (\mathcal{e}_{\nu} + \mathcal{P}_{\nu}) \hat{\bm u}_{\nu} \right]}{\partial \eta} + \nabla \cdot \left[ \gamma^2_{\nu} (\mathcal{e}_{\nu} + \mathcal{P}_{\nu}) \hat{\bm u}_{\nu} \hat{ \bm u}_{\nu} + \mathcal{P}_{\nu} \openone \right] \label{eq:fluid_neutrino_2} \\ 
		&= \sum_s q^W_{s\nu} \gamma_{\nu} \mathcal{n}_{\nu} \left( \mathbcal{E}_s + \hat{ \bm u}_{\nu} \times \mathbcal{B}_s \right),  \nonumber \\
		&\frac{ \partial \left[\gamma^2_{\nu} (\mathcal{e}_{\nu} + \mathcal{P}_{\nu} \hat{u}^2_{\nu}) \right]}{\partial \eta}  + \nabla \cdot \left[ \gamma^2_{\nu} (\mathcal{e}_{\nu} + \mathcal{P}_{\nu}) \hat{\bm u}_{\nu} \right] \label{eq:fluid_neutrino_3} \\ 
		&=  - \frac{a'}{a}  (3\mathcal{P}_{\nu} - \mathcal{e}_{\nu})  + \sum_s q^W_{s\nu} \gamma_{\nu} \mathcal{n}_{\nu}  \hat{ \bm u}_{\nu} \cdot \mathbcal{E}_s  , \nonumber
	\end{align}
	with the same continuity equation as Eq.~\eqref{eq:fluid_charged_1}. To close the system of fluid equations, an equation of state that relates internal energy and pressure is needed for each species; we will specify an appropriate equation of state for the fluids in Section~\ref{sec:NEP-MHD}-\ref{sec:NE-MHD}. As a last step, we express the effective fields of the weak and of the ponderomotive force in terms of conformal quantities. The procedure is straightforward and yields
	\begin{align}
		&\mathbcal{E}_{s,\nu} = - \frac{1}{a^2} \left[\nabla ( \gamma_{s,\nu} \mathcal{n}_{s,\nu}) +  \frac{\partial (\gamma_{s,\nu} \mathcal{n}_{s,\nu}  \hat{\bm u}_{s,\nu}) }{\partial \eta} \right. \label{eq:conformal_Ew}\\ 
		&\left. - \frac{3 a'}{a} \gamma_{s,\nu} \mathcal{n}_{s,\nu}  \hat{\bm u}_{s,\nu} \right], \quad \mathbcal{B}_{s,\nu} =  \frac{1}{a^2} \nabla \times (\gamma_{s,\nu} \mathcal{n}_{s,\nu}  \hat{\bm u}_{s,\nu}), \label{eq:conformal_Bw}
	\end{align}

	Note the presence of an additional term in Eq.~\eqref{eq:conformal_Ew} which is proportional to $a'/a = H a$, with $H$ the Hubble constant. This term originates from the conformal time derivative of the number current $\hat{\bm N}  = a^{-3} \gamma \mathcal{n} \hat{ \bm u} $ in the definition of the effective fields $\bm{E}_{s,\nu}, \bm{B}_{s,\nu}$ (Eq.~\eqref{eq:effective_weak_electric}-\eqref{eq:effective_weak_magnetic}) and can be understood as a correction due to the expansion of the Universe.
	
	\section{Simplified Fluid Models of Neutrino-Plasma Interactions}
	
	The equations derived thus far allow for a complete description of neutrino-plasma interactions in the FRW metric and in the hydrodynamic approximation, treating the particle species as separate fluids. Starting from the fluid equations, it is now possible to obtain simplified models that are suitable for analytical or numerical studies. In particular, we are interested in describing the effects of collective neutrino interactions on the generation of magnetic fields after the QCD transition and after the electron-positron annihilation.

	\subsection{Neutrino-Electron-Positron Plasma}\label{sec:NEP-MHD}
	
	At the QCD crossover ($T\sim 100 \,\, \si{MeV}$), the quarks and gluons combine to form baryons and mesons. Soon after the QCD crossover, nucleons begin to annihilate with their antiparticles, a process that is completed by the beginning of the lepton epoch, where the leptons dominated the mass content of the Universe \cite{Kolb1990}. As the number density of these heavier particles rapidly decreases, we can neglect their effect on the neutrinos and consider a pair plasma made of electrons and positrons.
	
	We assume that, while the electrons and the positrons are treated as a relativistic fluid, their bulk velocity satisfies $\hat{u}_e \ll c$. This is a reasonable approximation insofar as the individual particle motions are randomly distributed in all directions. For simplicity, we further assume that electrons and positrons have the same internal energy and pressure $ \mathcal{e}_e = \mathcal{e}_{\bar{e}} = \mathcal{e}/2$, $ \mathcal{P}_e = \mathcal{P}_{\bar{e}} = \mathcal{P}/2$, with $ \mathcal{e}, \mathcal{P}$ the internal energy and pressure of the electron-positron fluid.
	To describe the evolution of the electron-positron plasma, we introduce the total number density $\mathcal{n}$ and bulk fluid velocity $\hat{ \bm u}$
	\begin{align}
		\mathcal{n} = \mathcal{n}_{e} + \mathcal{n}_{\bar{e}}, \quad \hat{\bm u} = \frac{  \mathcal{n}_e  \hat{\bm u}_e + \mathcal{n}_{\bar{e}}  \hat{\bm u}_{\bar{e}}}{ \mathcal{n}_e  +  \mathcal{n}_{\bar{e}} }.
	\end{align}
	The electric current generated by the relative motion between the positive and negative charges is given by $\mathbcal{J} = e  (\mathcal{n}_{e} \hat{\bm u}_e - \mathcal{n}_{\bar{e}} \hat{\bm u}_{\bar{e}})$ and its evolution is obtained by combining the momentum equation ~\eqref{eq:fluid_charged_2} for the positrons and electrons in the so-called generalized Ohm's law
	\begin{widetext}
		\begin{align}
			&\frac{\partial }{\partial \eta} \left( \frac{e }{m_p}  (\mathcal{e}_p + \mathcal{P}_p) \hat{\bm u}_p - \frac{e }{m_e}  (\mathcal{e}_e + \mathcal{P}_e) \hat{\bm u}_e \right) + \nabla \cdot \left[  \frac{e }{m_p} (\mathcal{e}_p + \mathcal{P}_p) \hat{\bm u}_p \hat{\bm u}_p - \frac{e }{m_e}  (\mathcal{e}_e + \mathcal{P}_e) \hat{\bm u}_e \hat{\bm u}_e + \left( \frac{e \mathcal{P}_p}{m_p}  - \frac{e \mathcal{P}_e}{m_e}  \right)\delta_{ij} \right] \nonumber \\ 
			= e^2&\left( \frac{ \mathcal{n}_p }{m_p}  + \frac{ \mathcal{n}_e}{m_e}  \right) \mathbcal{E} + e^2 \left( \frac{ \mathcal{n}_p m_p + \mathcal{n}_e m_e}{m_p m_e}\right) \hat{\bm u} \times \mathbcal{B} - e\frac{m_p - m_e}{m_e m_p}  \mathbcal{J} \times \mathbcal{B} 
			+ \sum_{s=e,p} \sum_{\nu} q^W_{s\nu} \frac{ q_s  \mathcal{n}_s}{m_s} \left( \mathbcal{E}_{\nu} + \hat{\bm u}_s \times \mathbcal{B}_{\nu}\right),
			\label{eq:generalized_Ohm}
		\end{align}
	\end{widetext}
	where we have kept all the terms for later convenience.

	For an electron-positron plasma, Eq.\ref{eq:generalized_Ohm} simplifies considerably: in particular, we note that the Biermann battery terms cancel out following the assumption of equal pressure for the positrons and the electrons; moreover, the Hall term (proportional to $ \mathbcal{J} \times \mathbcal{B}$) also vanishes identically for an equal mass pair plasma. Finally, neglecting the time derivative and the nonlinear terms on the left-hand side -- as is customary in the derivation of the MHD equations -- the generalized Ohm's law reduces to
	\begin{align}
		\mathbcal{E} = -  \hat{\bm u} \times \mathbcal{B}    + \frac{ \sqrt{2}G_F  }{e } \left[ \left( \mathbcal{E}_{\nu} - \mathbcal{E}_{\bar{\nu}}  \right) \right. \label{eq:generalized_Ohm_ep_plasma} \\	\left. +  \hat{\bm u} \times \left( \mathbcal{B}_{\nu} - \mathbcal{B}_{\bar{\nu}}  \right) \right]. \nonumber
	\end{align}

	Following standard procedure, we obtain the fluid equations for the pair plasma combining Eqs.~\ref{eq:fluid_charged_1}-\ref{eq:fluid_charged_3} for the positrons and electrons under the assumption of quasi-neutrality ($\mathcal{n}_e = \mathcal{n}_{\bar{e}}$) and we use Eq.~\eqref{eq:generalized_Ohm_ep_plasma} to eliminate the electric field. Together with the induction equation, the fluid equations for the pair-neutrino plasma are
	\begin{widetext}
		\begin{align}
			&\frac{\partial \mathcal{n}}{\partial \eta} + \nabla \cdot \left( \mathcal{n} \hat{\bm u}\right) = 0, \label{eq:ep_1} \\
			&\frac{\partial \left[  (\mathcal{e} + \mathcal{P}) \hat{\bm u} \right]}{\partial \eta} + \nabla \cdot \left[  (\mathcal{e} + \mathcal{P}) \hat{\bm u} \hat{ \bm u} + \mathcal{P} \openone \right]  =  \mathbcal{ J} \times \mathbcal{B}  - \frac{\sqrt{2}G_F}{e } \mathbcal{ J} \times \left( \mathbcal{B}_{\nu} - \mathbcal{B}_{\bar{\nu}} \right)   ,  \label{eq:ep_2} \\
			&\frac{ \partial \mathcal{e} }{\partial \eta}  + \nabla \cdot \left[  (\mathcal{e} + \mathcal{P}) \hat{\bm u} \right]   = \mathbcal{ J} \cdot \mathbcal{ E} - \frac{\sqrt{2}G_F}{e }  \mathbcal{ J} \cdot \left( \mathbcal{E}_{\nu} - \mathbcal{E}_{\bar{\nu}} \right)   , \label{eq:ep_3}\\
			&\frac{\partial \mathbcal{B}}{\partial \eta} =  \nabla \times \left[   \hat{\bm u} \times  \mathbcal{B}  - \frac{\sqrt{2}G_F}{e } \left( \mathbcal{E}_{\nu} - \mathbcal{E}_{\bar{\nu}}  \right) -  \frac{\sqrt{2}G_F}{e } \hat{\bm u} \times \left( \mathbcal{B}_{\nu} - \mathbcal{B}_{\bar{\nu}}  \right) \right] , \label{eq:ep_4}
		\end{align}
	\end{widetext}
	with $\mathbcal{J}  = \nabla \times \mathbcal{B}/4\pi$, since the bulk motions of the plasma is nonrelativistic and thus the displacement current in Ampere's law can be neglected. Note, we implicitly use the ultra-relativistic equation of state $\mathcal{P}_{e,\bar{e}} = \mathcal{e}_{e,\bar{e}}/3$, and we have discarded the anisotropic terms in the pressure tensor.
	
	In the induction equation Eq.~\eqref{eq:ep_4} new terms have appeared due to the ponderomotive force exerted by the neutrinos on the electron-positron plasma (note that the electrostatic part of the effective potential, proportional to $\nabla \mathcal{n}_{\nu}$, vanishes identically in the curl). These terms are functions of the neutrino flux $\mathbcal{J}_{\nu} = \gamma_{\nu} \mathcal{n}_{\nu}  \hat{\bm u}_{\nu}$ and are independent of the value of $\mathbcal{B}$, and can thus act as a source for the generation of a seed magnetic field in the early Universe (see Section~\ref{sec:discussion} for further discussion).

	\subsubsection{Neutrino equations}
	
	The system of Eqs.~\ref{eq:ep_1}-\ref{eq:ep_4} is closed by the equations for the neutrinos, which we rewrite hereafter in terms of the neutrino flux $\mathbcal{J}_{\nu}$ and of the neutrino chemical potential $\mu_{\nu} \equiv (\mathcal{e}_{\nu} + \mathcal{P}_{\nu})/(\mathcal{n}_{\nu}a)$:
	\begin{align}
		&\frac{\partial (\gamma_{\nu} \mathcal{n}_{\nu} )}{\partial \eta} +\nabla \cdot \mathbcal{J}_{\nu}  = 0, \label{eq:neutrino_final_1} \\
		&\frac{\partial \left( a \gamma_{\nu}  \mu_{\nu}  \hat{\bm J}_{\nu} \right)}{\partial \eta} + \nabla \cdot \left[a  \gamma  \mu_{\nu} (\gamma_{\nu} \mathcal{n}_{\nu} )^{-1}  \hat{\bm J}_{\nu} \hat{ \bm J}_{\nu} + \mathcal{P}_{\nu} \openone \right] \nonumber  \\ 
		&=  \pm \sqrt{2}G_F  \left(  \gamma_{\nu} \mathcal{n}_{\nu} \left( \mathbcal{E}_e -  \mathbcal{E}_{\bar{e}} \right) +   \hat{\bm J}_{\nu} \times \left( \mathbcal{B}_e - \mathbcal{B}_{\bar{e}} \right) \right),  \label{eq:neutrino_final_2} \\
		&\frac{ \partial \left( a \gamma_{\nu}^2 \mu_{\nu} \mathcal{n}_{\nu} - \mathcal{P}_{\nu} \right)  }{\partial \eta}  + \nabla \cdot \left[ a \gamma_{\nu} \mu_{\nu}  \hat{\bm J}_{\nu} \right] =  - \frac{a'}{a}  \left(4\mathcal{P}_{\nu} \right.  \nonumber  \\ 
		& \left. - a \mu_{\nu} \mathcal{n}_{\nu}\right)   \pm \sqrt{2}G_F  \hat{ \bm J}_{\nu} \cdot \left( \mathbcal{E}_e -  \mathbcal{E}_{\bar{e}} \right) , \label{eq:neutrino_final_3}
	\end{align}
	where the subscript $\nu$ can take values of $\nu = \nu_e, \bar{\nu}_e$ and where the $+$($-$) sign refers to electron neutrinos (antineutrinos).
	
	For $T > m_{\nu}$, largely verified before the $e-p$ annihilation, the neutrinos are ultra-relativistic and the appropriate equation of state $\mathcal{P}_{\nu} = \mathcal{e}_{\nu}/3 = a \mu_{\nu} \mathcal{n}_{\nu}/4 $ can be used to further simplify the above equations. The opposite limit of cold neutrinos ($T < m_{\nu} $) can be obtained by neglecting thermal effects ($\mathcal{P}_{\nu} = 0$) and replacing $ \gamma_{\nu}  \mu_{\nu}  \rightarrow E_{\nu}$, where $E_{\nu}$ is the specific energy of the neutrino fluid. Finally, the equations of an ideal fluid consisting of massless neutrinos can be retrieved by neglecting the pressure and the trace of the stress-energy tensor (first term on the right-hand side of Eq.~\eqref{eq:neutrino_final_3}), and replacing $ \gamma_{\nu}  \mu_{\nu}  \rightarrow E_{\nu}$ throughout.

%
%
%
%
	
	\subsection{Neutrino-Electron MHD}\label{sec:NE-MHD}
	
	After the electron-positron annihilation ($T < 0.5 \,\, \si{MeV}$), only a small excess of electrons survived and the global charge of the Universe resides in the electrons and in the protons \cite{Kolb1990}. These particles interact with the relativistic neutrinos and their evolution can be appropriately described through a magneto-hydrodynamic formulation that treats the protons and electrons as a single-fluid, while the neutrinos are evolved separately, as done in \cite{Haas2016} for neutrino-plasma interactions in a flat metric (see also \cite{Gailis1995,Brandenburg1996} for MHD models in an FRW metric but without neutrinos). A simpler approach, which we adopt in this work, is to neglect the motion of the protons and develop a model to study the dynamics of electrons and neutrinos only. This framework, that we call neutrino-electron MHD (NEMHD), is similar in spirit to electron MHD (EMHD) models \cite{Lighthill1960,Kingsep1987,Gordeev1994,Biskampd1999}, where the ions only form a neutralizing background and the electron inertia is neglected. Note that, while in this section we mostly focus on electron-proton plasmas, we will use the terms "ions" and "protons" interchangeably to account for the fact that -- as the temperature decreases -- the plasma will also contain heavier elements.\\
	
	To derive the NEMHD model, we start from the generalized Ohm's law assuming nonrelativistic bulk motion of the ions and electrons. The derivation is analogous to the previous case of the electron-positron plasma and the final result is formally identical to Eq.~\eqref{eq:generalized_Ohm} with the replacement of the subscript $p$ (for positrons) by $i$ (for ions), and where the ion-electron bulk velocity $\hat{\bm u}$ and current density $\mathbcal{J}$ now are
	\begin{align}
		\hat{\bm u} = \frac{ \mathcal{n}_i m_i \hat{\bm u}_i +  \mathcal{n}_e m_e \hat{\bm u}_e}{ \mathcal{n}_i m_i  +  \mathcal{n}_e m_e }, \quad \mathbcal{J} = e \left(  \mathcal{n}_i \hat{\bm u}_i -  \mathcal{n}_e \hat{\bm u}_e \right).
	\end{align}
	
	In order to simplify Eq.~\eqref{eq:generalized_Ohm}, we assume plasma quasineutrality ($ \mathcal{n}_i \simeq  \mathcal{n}_e$) and then take the limit of slow ions $\hat{ \bm u}_i \ll \hat{ \bm u}_e$ and inertialess electrons. These last two assumptions define the EMHD approximation, that is applicable on length-scales ($\ell$) in the range $c / \omega_{pe} \ll \ell \ll c / \omega_{pi}$, with $\omega_{pe,i} = (4 \pi n_{e,i} e^2 / \gamma_{e,i} m_{e,i} )^{1/2}$ the electron (ion) plasma frequency \cite{Kingsep1987,Gordeev1994}. In this regime, the second-order terms in the divergence and the time derivative of the generalized Ohm's law proportional to the electron velocity can safely be neglected, as they would introduce a correction of order $\mathcal{O}(c^2/\omega_{pe}^2 \ell^2)$ in the induction equation. With these assumptions, $\hat{\bm u} \simeq \hat{\bm u}_i$, $\mathbcal{J} \simeq - e \gamma_e \mathcal{n}_e \hat{\bm u}_e $, and the generalized Ohm's law simplifies to 
	\begin{align}\label{eq:simpl_Ohm}
		\mathbcal{E} = -\frac{ \nabla \mathcal{P}_e}{e  \mathcal{n}_{e}} &+  \frac{\left( \nabla \times  \mathbcal{B} \right) \times  \mathbcal{B} }{4 \pi e  \mathcal{n}_{e}}  \\
		&+ \sum_{\nu = \nu_e, \bar{\nu}_e} \frac{q^W_{e\nu}}{e} \left( \mathbcal{E}_{\nu} - \frac{\left( \nabla \times  \mathbcal{B} \right)}{4 \pi e  \mathcal{n}_e} \times \mathbcal{B}_{\nu} \right),  \nonumber 
	\end{align}
	where only electron-neutrinos and electron-antineutrinos now contribute to the generation of electric fields as a result of the EMHD approximation. The limit of slow ions and inertialess electrons has two further important consequences: (i) the dynamics of the ions is effectively decoupled from that of electrons, as they only provide a neutralizing background for the electron flow, and (ii) the motion of the electrons is not independent but follows from the evolution of the magnetic field through the induction equation, since $\hat{ \bm u}_e = -(4 \pi e \mathcal{n}_e )^{-1} \nabla \times \mathbcal{B} $. In fact, replacing $\hat{ \bm u}_e$ in the continuity equation Eq.~(\ref{eq:fluid_charged_1}), we find that $ \mathcal{n}_e$ remains constant in conformal time. As a result, the internal dynamics of the electrons is entirely determined by the induction equation Eq.~\eqref{eq:max_frw_2} and by the internal energy equation, which we rewrite using Eq.~\eqref{eq:simpl_Ohm} in the limit of a non-relativistic gas as follows
	\begin{align}
		\frac{ \partial  \left( a  \mathcal{P}_e\right) }{\partial \eta}  + \hat{\bm u}_e \cdot \nabla \left(  a \mathcal{P}_e \right)   + \frac{5}{3} a \mathcal{P}_e \nabla \cdot \hat{ \bm u}_e  = 0,  \label{eq:electron_int_energy} 
	\end{align}
	where we used the adiabatic equation of state for the electron gas. Note that Eq.~\eqref{eq:electron_int_energy} is the conformal counterpart of the energy equation of an ideal fluid with adiabatic index equal to $5/3$, and is equivalent to stating that the (modified) entropy density of the electrons $\mathcal{s}_e = \ln a \mathcal{P}_e \mathcal{n}_e^{-5/3}$ is a material invariant. In the same non-relativistic limit for the electrons, the leading contribution (up to order $\mathcal{O}(\hat{u}_e^2/c^2)$) to the effective fields of the weak interaction Eq.~\eqref{eq:conformal_Ew}-\eqref{eq:conformal_Bw} is
	\begin{align}
		\mathbcal{E}_e \simeq - \frac{1}{a^2} \nabla  \mathcal{n}_e , \quad \mathbcal{B}_e \simeq  0. \label{eq:conformal_Ew_Bw_nonrel}
	\end{align}
	We remark that, in the case of a positron-electron plasma analyzed before, the zeroth order contributions of the positrons and the electrons to the neutrino equations cancel out, and thus it was necessary to keep the higher-order corrections to accurately describe the neutrino-plasma interaction. For an electron-proton plasma no such cancellation occurs and the relativistic corrections can be neglected.
	
	We are now in position to write the full NEMHD system for neutrino-electron interaction, that
	consists of the following equations:
	%
	\begin{widetext}
		\begin{align}
			&\frac{ \partial  \mathcal{s}_e}{\partial \eta}  = \frac{ \nabla \times  \mathbcal{B} }{4 \pi e \mathcal{n}_{e}} \cdot \nabla  \mathcal{s}_e  ,  \label{eq:final_eqs_electron}  \\
			&\frac{\partial (\gamma_{\nu} \mathcal{n}_{\nu} )}{\partial \eta} +\nabla \cdot \mathbcal{J}_{\nu}  = 0, \label{eq:neutrino_nemhd_1} \\
			&\frac{\partial \left( a \gamma_{\nu}  \mu_{\nu}  \hat{\bm J}_{\nu} \right)}{\partial \eta} + \nabla \cdot \left[a  \gamma  \mu_{\nu} (\gamma_{\nu} \mathcal{n}_{\nu} )^{-1}  \hat{\bm J}_{\nu} \hat{ \bm J}_{\nu} + \mathcal{P}_{\nu} \openone \right] =  \mp \frac{\sqrt{2}G_F  \gamma_{\nu} \mathcal{n}_{\nu}}{a^2}   \nabla  \mathcal{n}_{e} ,  \label{eq:neutrino_nemhd_2} \\
			&\frac{ \partial \left( a \gamma_{\nu}^2 \mu_{\nu} \mathcal{n}_{\nu} - \mathcal{P}_{\nu} \right)  }{\partial \eta}  + \nabla \cdot \left[ a \gamma_{\nu} \mu_{\nu}  \hat{\bm J}_{\nu} \right] =  - \frac{a'}{a}  \left(4\mathcal{P}_{\nu}  - a \mu_{\nu} \mathcal{n}_{\nu}\right)   \mp \frac{\sqrt{2}G_F}{a^2}  \hat{ \bm J}_{\nu} \cdot  \nabla  \mathcal{n}_{e}  , \label{eq:neutrino_nemhd_3} \\
			&\frac{\partial \mathbcal{B}}{\partial \eta} =  \nabla \times \left[ \frac{ \nabla \mathcal{P}_e}{e \mathcal{n}_{e}} -  \frac{\left( \nabla \times  \mathbcal{B} \right) \times  \mathbcal{B} }{4 \pi e  \mathcal{n}_{e}}  +  \frac{\sqrt{2}G_F}{e a^2} \left( \frac{\partial}{\partial \eta}  - \frac{3 a'}{a} \right) \left( \mathbcal{J}_{\nu} - \mathbcal{J}_{\bar{\nu}} \right) + \frac{\sqrt{2}G_F}{e a^2} \frac{\left( \nabla \times  \mathbcal{B} \right)}{4 \pi e  \mathcal{n}_e} \times \nabla \times \left( \mathbcal{J}_{\nu} - \mathbcal{J}_{\bar{\nu}} \right) \right] , \label{eq:final_eqs_induction}   
		\end{align}
	\end{widetext}
	where we expressed the energy equation of the electrons in terms of the specific entropy, and where the $-$($+$) sign now refers to electron neutrinos (antineutrinos). Note that in the NEMHD equations $\mathcal{n}_e$ is not a dynamical variable, but is constant in time (though it may vary in space), as a result of the electron-MHD approximation.

	\section{Discussion}\label{sec:discussion}
	An important application of collective neutrino-plasma interactions is the self-consistent production (or the subsequent amplification) of a primordial magnetic seed. In the early universe, magnetic fields can potentially be generated through a variety of processes, including, but not limited to, inflationary production \cite{Turner1988}, first-order phase transitions \cite{Sigl1997} -- where the Biermann battery \cite{Biermann1950} is also expected to operate  -- and plasma vorticity in the late radiation era \cite{Harrison1970} (see, e.g. \cite{Grasso2001,Durrer2013} for reviews).

	In our model, magnetic fields can be generated if the difference between the neutrino and antineutrino currents is nonzero, as we can see from Eq.~(\ref{eq:ep_4}) and (\ref{eq:final_eqs_induction}). In fact, if we take Eq.~(\ref{eq:neutrino_final_2}) and subtract the corresponding equation for antineutrinos, the terms proportional to the effective fields $\mathbcal{E}_e$ and $\mathbcal{B}_e$ do not cancel out -- thanks to Eq.~(\ref{eq:charge_identities}) -- but rather add up to produce a difference in the neutrino fluxes. From a physical point of view, this is the result of the presence of inhomogeneities in the electron distribution that push the neutrinos and the antineutrinos away from each other.

	In an electron-positron plasma, however, the contributions of electrons and positrons act in opposite directions and self-consistent generation of a seed magnetic field is possible only in presence of a local net charge imbalance $\mathcal{n}_e - \mathcal{n}_{\bar{e}} \neq 0 $. This is in fact the case in our universe, where at the time of the QCD crossover there was a small excess of electrons compared to positrons of the order of $10^{-9}$ \cite{Langacker1983}, which resulted in the presence of leftover electrons at present-day. In such a scenario, magnetic fields can be generated through a "neutrino battery" mechanism (in clear reference to the well-known Biermann battery), whereby, substituting $\partial \left( \mathbcal{J}_{\nu} - \mathbcal{J}_{\bar{\nu}} \right) / \partial \eta$  in
	the equation for the electric field, there appears the following new term
	\begin{align}
		\mathbcal{E} = ... - \frac{2 G_F^2 }{e a^3 k_B T } (\mathcal{n}_{\nu} +  \mathcal{n}_{\bar{\nu}} ) \nabla ( \mathcal{n}_e - \mathcal{n}_{\bar{e}})
	\end{align} 
	(where we assumed for simplicity that neutrinos and antineutrinos have non-relativistic bulk motions and the same energy $E_{\nu} \approx k_B T$), which has a rotational component if neutrino gradients are misaligned with electron/positron gradients, and is present even in the case of an initially zero neutrino flux difference. Contrary to the usual Biermann term, the neutrino battery does not vanish if the plasma remains barotropic during the QCD crossover (as one would expect if the QCD is not, in fact, a first-order phase transition \cite{Aoki2006}), since the turbulent fluctuations in the thermodynamic properties of the plasma generated at the phase transition will not be perfectly correlated with those of the neutrino field. The relevance of this result also lies in the fact that it does not require physics outside of the Standard Model to explain primordial magnetogenesis, such as axion-photon coupling \cite{Miniati2018}, or other processes that break conformal invariance (see, e.g., \cite{Durrer2013} and references therein).

	Assuming a lepton asymmetry of the order of $\sim 10^{-9} \mathcal{n}_e$, we can estimate the electric field produced by the neutrinos as
	\begin{align}
		eE \simeq 10^{-9} a^2  \left( \frac{k_B T}{(\sqrt{2}G_F^0)^{-1/2}}\right)^4 \frac{k_B T}{L_H/2} ,	
	\end{align} 
	where $L_H$ is the particle horizon, which generates a magnetic seed  at the QCD crossover of $B_{QCD} \sim 10^{-46} \si{G}$. Despite being very small, this initial seed can undergo significant amplification before being damped by cosmological expansion through a small-scale dynamo produced by the turbulence at the QCD. In fact, as suggested by lattice simulations \cite{Ejiri2006,Bazavov2013,Bellwied2015}, if turbulent velocity fluctuations of the order of $\delta u \sim 1/\sqrt{3 g^*}$ (with $g^*$ the effective number of degrees of freedom) are excited on scales of $l \sim 0.1 L_H$, the Reynolds number at the crossover is expected to be large ($Re \simeq \delta u \times l / \lambda_{\mathrm{mfp},\nu} \gtrsim 10^4$, where $\lambda_{\mathrm{mfp},\nu}$ is the neutrino mean-free path), and the magnetic seed grows exponentially on a timescale comparable to the viscous-scale eddy turnover time $\sim Re^{-1/2} t_{QCD} \ll t_{QCD}$, reaching the equipartition field strength (in comoving units) with the turbulent energy of $\sim \si{\mu G}$  \cite{Miniati2018}. If the magnetic field is then frozen-in in the expanding plasma, this equipartition value would correspond to a magnetic field strength at recombination of $\sim 10^{-3} \si{nG}$ \cite{Miniati2018}, close to the $5\times 10^{-3} - 0.1 \si{nG}$ range which results from constraints from the cosmic microwave background anisotropy and current magnetic fields in galaxy clusters \cite{Banerjee2004,Trivedi2010,Jedamzik2019,Jedamzik2020}. We remark that the effectiveness of the small-scale dynamo in amplifying the small magnetic seed strongly depends on the turbulent levels at the QCD crossover. Estimates of the $\delta u$ based on primordial density fluctuations would put the plasma at the QCD transition in the highly-subsonic regime, reducing the Reynolds number and slowing down the magnetic seed amplification \cite{Wagstaff2014,Chirakkara2021}. 

	
%
%

	The neutrino battery mechanism as outlined above can be understood as a second-order process (i.e. proportional to $G_F^2$), whereby small inhomogeneities in the electron/positron distribution create a non-zero net neutrino-flux, which in turn generates an electric field.
	Alternatively, if a net neutrino-flux is already present at the QCD phase transition (e.g., as a result of the turbulence at the crossover), magnetic fields can be generated and amplified through such terms as $ a'/a ( \mathbcal{J}_{\nu} - \mathbcal{J}_{\bar{\nu}} )$ and $ \nabla \times ( \mathbcal{J}_{\nu} - \mathbcal{J}_{\bar{\nu}} )$ that are proportional to $G_F$ and therefore constitute first-order processes. A similar distinction was also made in \cite{Brizard2000}.
	
	We note that the presence of inhomogeneities in the neutrino distribution are not strictly required for the production of magnetic field. In fact, differences between the number densities of neutrinos and antineutrinos are sufficient to generate strong magnetic fields. This is likely to be the case, e.g., in the core of proto-neutron stars, where electron neutrinos are produced in large numbers by electron capture on nucleons \cite{Burrows1990,Janka2017}, or in the early universe in presence of a neutrino asymmetry \cite{Lesgourgues1999,Dolgov2002b}.

	\subsection{Summary and conclusions}
	
	In this paper, we derived a theoretical framework to study the effect of collective interactions between neutrinos and the plasma in an expanding universe.
	Starting from the relativistic kinetic equations for the particle distribution function in the FRW metric, we obtained a simplified fluid model that attempts to capture the main effects of neutrino-plasma interactions. In particular, we looked at two different scenarios where the neutrinos could lead to generation of a primordial magnetic field, namely (a) the lepton epoch, where we consider a pair electron-positron plasma (Section~\ref{sec:NEP-MHD}), and (b) at the end of electron-positron annihilation (Section~\ref{sec:NE-MHD}), where we look at an electron-proton plasma in the limit of slow ions and inertialess electrons. In both scenarios, we have identified a promising mechanism that can generate primordial magnetic fields based on a "neutrino battery" process, whereby misaligned gradients in the number density of the neutrino and electron populations act as a source term in the induction equation.
	
	Our model differs from that of \cite{Brizard2000}, who derived the relativistic fluid equations for collective neutrino-plasma interactions through a Lagrangian variational principle, in that we focused our attention on the 
	effects of an expanding space-time on the neutrino-plasma dynamics, and found that new terms appear in the induction equation that are proportional to the expansion rate of the universe. To our knowledge, this result had not been obtained in the literature before. On the other hand, for a static universe ($a = 1$), Equations ~\eqref{eq:fluid_charged_1}-\eqref{eq:fluid_neutrino_3} reduce to those in \cite{Brizard2000}. Due to their simplicity, our equations can thus serve as the basis for further numerical or analytical studies of magnetic field generation in the early universe. A more detailed discussion on the potential applications of our model is left for future work.

	\begin{acknowledgments}
		We wish to thank the anonymous referee for their constructive feedback, that lead to a significant improvement of the paper.
	    L.M.P. is funded by the STFC CDT in Data Intensive Science at the University of Cambridge, and acknowledges the hospitality of the Rutherford Appleton Laboratory and of the Atomic and Laser Physics Department at the University of Oxford where this work was carried out as part of an internship. The research leading to these results have been funded in parts by the
		Engineering and Physical Sciences Research Council
		(grant numbers EP/M022331/1 and EP/N014472/1).
	\end{acknowledgments}
	
	\bibliography{bibliography.bib}

\begin{thebibliography}{71}%
\makeatletter
\providecommand \@ifxundefined [1]{%
 \@ifx{#1\undefined}
}%
\providecommand \@ifnum [1]{%
 \ifnum #1\expandafter \@firstoftwo
 \else \expandafter \@secondoftwo
 \fi
}%
\providecommand \@ifx [1]{%
 \ifx #1\expandafter \@firstoftwo
 \else \expandafter \@secondoftwo
 \fi
}%
\providecommand \natexlab [1]{#1}%
\providecommand \enquote  [1]{``#1''}%
\providecommand \bibnamefont  [1]{#1}%
\providecommand \bibfnamefont [1]{#1}%
\providecommand \citenamefont [1]{#1}%
\providecommand \href@noop [0]{\@secondoftwo}%
\providecommand \href [0]{\begingroup \@sanitize@url \@href}%
\providecommand \@href[1]{\@@startlink{#1}\@@href}%
\providecommand \@@href[1]{\endgroup#1\@@endlink}%
\providecommand \@sanitize@url [0]{\catcode `\\12\catcode `\$12\catcode
  `\&12\catcode `\#12\catcode `\^12\catcode `\_12\catcode `\%12\relax}%
\providecommand \@@startlink[1]{}%
\providecommand \@@endlink[0]{}%
\providecommand \url  [0]{\begingroup\@sanitize@url \@url }%
\providecommand \@url [1]{\endgroup\@href {#1}{\urlprefix }}%
\providecommand \urlprefix  [0]{URL }%
\providecommand \Eprint [0]{\href }%
\providecommand \doibase [0]{https://doi.org/}%
\providecommand \selectlanguage [0]{\@gobble}%
\providecommand \bibinfo  [0]{\@secondoftwo}%
\providecommand \bibfield  [0]{\@secondoftwo}%
\providecommand \translation [1]{[#1]}%
\providecommand \BibitemOpen [0]{}%
\providecommand \bibitemStop [0]{}%
\providecommand \bibitemNoStop [0]{.\EOS\space}%
\providecommand \EOS [0]{\spacefactor3000\relax}%
\providecommand \BibitemShut  [1]{\csname bibitem#1\endcsname}%
\let\auto@bib@innerbib\@empty
\bibitem [{\citenamefont {{Bahcall}}(1989)}]{Bahcall1989}%
  \BibitemOpen
  \bibfield  {author} {\bibinfo {author} {\bibfnamefont {J.~N.}\ \bibnamefont
  {{Bahcall}}},\ }\href@noop {} {\emph {\bibinfo {title} {{Neutrino
  Astrophysics}}}}\ (\bibinfo {year} {1989})\BibitemShut {NoStop}%
\bibitem [{\citenamefont {{Bionta}}\ \emph {et~al.}(1987)\citenamefont
  {{Bionta}}, \citenamefont {{Blewitt}}, \citenamefont {{Bratton}},
  \citenamefont {{Casper}}, \citenamefont {{Ciocio}}, \citenamefont {{Claus}},
  \citenamefont {{Cortez}}, \citenamefont {{Crouch}}, \citenamefont {{Dye}},
  \citenamefont {{Errede}}, \citenamefont {{Foster}}, \citenamefont
  {{Gajewski}}, \citenamefont {{Ganezer}}, \citenamefont {{Goldhaber}},
  \citenamefont {{Haines}}, \citenamefont {{Jones}}, \citenamefont
  {{Kielczewska}}, \citenamefont {{Kropp}}, \citenamefont {{Learned}},
  \citenamefont {{Losecco}}, \citenamefont {{Matthews}}, \citenamefont
  {{Miller}}, \citenamefont {{Mudan}}, \citenamefont {{Park}}, \citenamefont
  {{Price}}, \citenamefont {{Reines}}, \citenamefont {{Schultz}}, \citenamefont
  {{Seidel}}, \citenamefont {{Shumard}}, \citenamefont {{Sinclair}},
  \citenamefont {{Sobel}}, \citenamefont {{Stone}}, \citenamefont {{Sulak}},
  \citenamefont {{Svoboda}}, \citenamefont {{Thornton}}, \citenamefont {{van
  der Velde}},\ and\ \citenamefont {{Wuest}}}]{Bionta1987}%
  \BibitemOpen
  \bibfield  {author} {\bibinfo {author} {\bibfnamefont {R.~M.}\ \bibnamefont
  {{Bionta}}}, \bibinfo {author} {\bibfnamefont {G.}~\bibnamefont {{Blewitt}}},
  \bibinfo {author} {\bibfnamefont {C.~B.}\ \bibnamefont {{Bratton}}}, \bibinfo
  {author} {\bibfnamefont {D.}~\bibnamefont {{Casper}}}, \bibinfo {author}
  {\bibfnamefont {A.}~\bibnamefont {{Ciocio}}}, \bibinfo {author}
  {\bibfnamefont {R.}~\bibnamefont {{Claus}}}, \bibinfo {author} {\bibfnamefont
  {B.}~\bibnamefont {{Cortez}}}, \bibinfo {author} {\bibfnamefont
  {M.}~\bibnamefont {{Crouch}}}, \bibinfo {author} {\bibfnamefont {S.~T.}\
  \bibnamefont {{Dye}}}, \bibinfo {author} {\bibfnamefont {S.}~\bibnamefont
  {{Errede}}}, \bibinfo {author} {\bibfnamefont {G.~W.}\ \bibnamefont
  {{Foster}}}, \bibinfo {author} {\bibfnamefont {W.}~\bibnamefont
  {{Gajewski}}}, \bibinfo {author} {\bibfnamefont {K.~S.}\ \bibnamefont
  {{Ganezer}}}, \bibinfo {author} {\bibfnamefont {M.}~\bibnamefont
  {{Goldhaber}}}, \bibinfo {author} {\bibfnamefont {T.~J.}\ \bibnamefont
  {{Haines}}}, \bibinfo {author} {\bibfnamefont {T.~W.}\ \bibnamefont
  {{Jones}}}, \bibinfo {author} {\bibfnamefont {D.}~\bibnamefont
  {{Kielczewska}}}, \bibinfo {author} {\bibfnamefont {W.~R.}\ \bibnamefont
  {{Kropp}}}, \bibinfo {author} {\bibfnamefont {J.~G.}\ \bibnamefont
  {{Learned}}}, \bibinfo {author} {\bibfnamefont {J.~M.}\ \bibnamefont
  {{Losecco}}}, \bibinfo {author} {\bibfnamefont {J.}~\bibnamefont
  {{Matthews}}}, \bibinfo {author} {\bibfnamefont {R.}~\bibnamefont
  {{Miller}}}, \bibinfo {author} {\bibfnamefont {M.~S.}\ \bibnamefont
  {{Mudan}}}, \bibinfo {author} {\bibfnamefont {H.~S.}\ \bibnamefont {{Park}}},
  \bibinfo {author} {\bibfnamefont {L.~R.}\ \bibnamefont {{Price}}}, \bibinfo
  {author} {\bibfnamefont {F.}~\bibnamefont {{Reines}}}, \bibinfo {author}
  {\bibfnamefont {J.}~\bibnamefont {{Schultz}}}, \bibinfo {author}
  {\bibfnamefont {S.}~\bibnamefont {{Seidel}}}, \bibinfo {author}
  {\bibfnamefont {E.}~\bibnamefont {{Shumard}}}, \bibinfo {author}
  {\bibfnamefont {D.}~\bibnamefont {{Sinclair}}}, \bibinfo {author}
  {\bibfnamefont {H.~W.}\ \bibnamefont {{Sobel}}}, \bibinfo {author}
  {\bibfnamefont {J.~L.}\ \bibnamefont {{Stone}}}, \bibinfo {author}
  {\bibfnamefont {L.~R.}\ \bibnamefont {{Sulak}}}, \bibinfo {author}
  {\bibfnamefont {R.}~\bibnamefont {{Svoboda}}}, \bibinfo {author}
  {\bibfnamefont {G.}~\bibnamefont {{Thornton}}}, \bibinfo {author}
  {\bibfnamefont {J.~C.}\ \bibnamefont {{van der Velde}}},\ and\ \bibinfo
  {author} {\bibfnamefont {C.}~\bibnamefont {{Wuest}}},\ }\bibfield  {title}
  {\bibinfo {title} {{Observation of a neutrino burst in coincidence with
  supernova 1987A in the Large Magellanic Cloud}},\ }\href
  {https://doi.org/10.1103/PhysRevLett.58.1494} {\bibfield  {journal} {\bibinfo
   {journal} {\prl}\ }\textbf {\bibinfo {volume} {58}},\ \bibinfo {pages}
  {1494} (\bibinfo {year} {1987})}\BibitemShut {NoStop}%
\bibitem [{\citenamefont {{Hirata}}\ \emph {et~al.}(1987)\citenamefont
  {{Hirata}}, \citenamefont {{Kajita}}, \citenamefont {{Koshiba}},
  \citenamefont {{Nakahata}}, \citenamefont {{Oyama}}, \citenamefont {{Sato}},
  \citenamefont {{Suzuki}}, \citenamefont {{Takita}}, \citenamefont
  {{Totsuka}}, \citenamefont {{Kifune}}, \citenamefont {{Suda}}, \citenamefont
  {{Takahashi}}, \citenamefont {{Tanimori}}, \citenamefont {{Miyano}},
  \citenamefont {{Yamada}}, \citenamefont {{Beier}}, \citenamefont
  {{Feldscher}}, \citenamefont {{Kim}}, \citenamefont {{Mann}}, \citenamefont
  {{Newcomer}}, \citenamefont {{van}}, \citenamefont {{Zhang}},\ and\
  \citenamefont {{Cortez}}}]{Hirata1987}%
  \BibitemOpen
  \bibfield  {author} {\bibinfo {author} {\bibfnamefont {K.}~\bibnamefont
  {{Hirata}}}, \bibinfo {author} {\bibfnamefont {T.}~\bibnamefont {{Kajita}}},
  \bibinfo {author} {\bibfnamefont {M.}~\bibnamefont {{Koshiba}}}, \bibinfo
  {author} {\bibfnamefont {M.}~\bibnamefont {{Nakahata}}}, \bibinfo {author}
  {\bibfnamefont {Y.}~\bibnamefont {{Oyama}}}, \bibinfo {author} {\bibfnamefont
  {N.}~\bibnamefont {{Sato}}}, \bibinfo {author} {\bibfnamefont
  {A.}~\bibnamefont {{Suzuki}}}, \bibinfo {author} {\bibfnamefont
  {M.}~\bibnamefont {{Takita}}}, \bibinfo {author} {\bibfnamefont
  {Y.}~\bibnamefont {{Totsuka}}}, \bibinfo {author} {\bibfnamefont
  {T.}~\bibnamefont {{Kifune}}}, \bibinfo {author} {\bibfnamefont
  {T.}~\bibnamefont {{Suda}}}, \bibinfo {author} {\bibfnamefont
  {K.}~\bibnamefont {{Takahashi}}}, \bibinfo {author} {\bibfnamefont
  {T.}~\bibnamefont {{Tanimori}}}, \bibinfo {author} {\bibfnamefont
  {K.}~\bibnamefont {{Miyano}}}, \bibinfo {author} {\bibfnamefont
  {M.}~\bibnamefont {{Yamada}}}, \bibinfo {author} {\bibfnamefont {E.~W.}\
  \bibnamefont {{Beier}}}, \bibinfo {author} {\bibfnamefont {L.~R.}\
  \bibnamefont {{Feldscher}}}, \bibinfo {author} {\bibfnamefont {S.~B.}\
  \bibnamefont {{Kim}}}, \bibinfo {author} {\bibfnamefont {A.~K.}\ \bibnamefont
  {{Mann}}}, \bibinfo {author} {\bibfnamefont {F.~M.}\ \bibnamefont
  {{Newcomer}}}, \bibinfo {author} {\bibfnamefont {R.}~\bibnamefont {{van}}},
  \bibinfo {author} {\bibfnamefont {W.}~\bibnamefont {{Zhang}}},\ and\ \bibinfo
  {author} {\bibfnamefont {B.~G.}\ \bibnamefont {{Cortez}}},\ }\bibfield
  {title} {\bibinfo {title} {{Observation of a neutrino burst from the
  supernova SN1987A}},\ }\href {https://doi.org/10.1103/PhysRevLett.58.1490}
  {\bibfield  {journal} {\bibinfo  {journal} {\prl}\ }\textbf {\bibinfo
  {volume} {58}},\ \bibinfo {pages} {1490} (\bibinfo {year}
  {1987})}\BibitemShut {NoStop}%
\bibitem [{\citenamefont {{Bahcall}}\ and\ \citenamefont
  {{Sears}}(1972)}]{Bahcall1972}%
  \BibitemOpen
  \bibfield  {author} {\bibinfo {author} {\bibfnamefont {J.~N.}\ \bibnamefont
  {{Bahcall}}}\ and\ \bibinfo {author} {\bibfnamefont {R.~L.}\ \bibnamefont
  {{Sears}}},\ }\bibfield  {title} {\bibinfo {title} {{Solar Neutrinos}},\
  }\href {https://doi.org/10.1146/annurev.aa.10.090172.000325} {\bibfield
  {journal} {\bibinfo  {journal} {\araa}\ }\textbf {\bibinfo {volume} {10}},\
  \bibinfo {pages} {25} (\bibinfo {year} {1972})}\BibitemShut {NoStop}%
\bibitem [{\citenamefont {{Steigman}}(2012)}]{Steigman2012}%
  \BibitemOpen
  \bibfield  {author} {\bibinfo {author} {\bibfnamefont {G.}~\bibnamefont
  {{Steigman}}},\ }\bibfield  {title} {\bibinfo {title} {{Neutrinos And Big
  Bang Nucleosynthesis}},\ }\href@noop {} {\bibfield  {journal} {\bibinfo
  {journal} {arXiv e-prints}\ ,\ \bibinfo {eid} {arXiv:1208.0032}} (\bibinfo
  {year} {2012})},\ \Eprint {https://arxiv.org/abs/1208.0032} {arXiv:1208.0032
  [hep-ph]} \BibitemShut {NoStop}%
\bibitem [{\citenamefont {{Bingham}}\ \emph {et~al.}(1994)\citenamefont
  {{Bingham}}, \citenamefont {{Dawson}}, \citenamefont {{Su}},\ and\
  \citenamefont {{Bethe}}}]{Bingham1994}%
  \BibitemOpen
  \bibfield  {author} {\bibinfo {author} {\bibfnamefont {R.}~\bibnamefont
  {{Bingham}}}, \bibinfo {author} {\bibfnamefont {J.~M.}\ \bibnamefont
  {{Dawson}}}, \bibinfo {author} {\bibfnamefont {J.~J.}\ \bibnamefont {{Su}}},\
  and\ \bibinfo {author} {\bibfnamefont {H.~A.}\ \bibnamefont {{Bethe}}},\
  }\bibfield  {title} {\bibinfo {title} {{Collective interactions between
  neutrinos and dense plasmas}},\ }\href
  {https://doi.org/10.1016/0375-9601(94)90597-5} {\bibfield  {journal}
  {\bibinfo  {journal} {Physics Letters A}\ }\textbf {\bibinfo {volume}
  {193}},\ \bibinfo {pages} {279} (\bibinfo {year} {1994})}\BibitemShut
  {NoStop}%
\bibitem [{\citenamefont {Bingham}\ \emph {et~al.}(1996)\citenamefont
  {Bingham}, \citenamefont {Bethe}, \citenamefont {Dawson}, \citenamefont
  {Shukla},\ and\ \citenamefont {Su}}]{Bingham1996}%
  \BibitemOpen
  \bibfield  {author} {\bibinfo {author} {\bibfnamefont {R.}~\bibnamefont
  {Bingham}}, \bibinfo {author} {\bibfnamefont {H.}~\bibnamefont {Bethe}},
  \bibinfo {author} {\bibfnamefont {J.}~\bibnamefont {Dawson}}, \bibinfo
  {author} {\bibfnamefont {P.}~\bibnamefont {Shukla}},\ and\ \bibinfo {author}
  {\bibfnamefont {J.}~\bibnamefont {Su}},\ }\bibfield  {title} {\bibinfo
  {title} {Nonlinear scattering of neutrinos by plasma waves: a ponderomotive
  force description},\ }\href
  {https://doi.org/https://doi.org/10.1016/0375-9601(96)00503-8} {\bibfield
  {journal} {\bibinfo  {journal} {Physics Letters A}\ }\textbf {\bibinfo
  {volume} {220}},\ \bibinfo {pages} {107} (\bibinfo {year}
  {1996})}\BibitemShut {NoStop}%
\bibitem [{\citenamefont {{Silva}}\ \emph
  {et~al.}(1999{\natexlab{a}})\citenamefont {{Silva}}, \citenamefont
  {{Bingham}}, \citenamefont {{Dawson}},\ and\ \citenamefont
  {{Mori}}}]{Silva1999E}%
  \BibitemOpen
  \bibfield  {author} {\bibinfo {author} {\bibfnamefont {L.~O.}\ \bibnamefont
  {{Silva}}}, \bibinfo {author} {\bibfnamefont {R.}~\bibnamefont {{Bingham}}},
  \bibinfo {author} {\bibfnamefont {J.~M.}\ \bibnamefont {{Dawson}}},\ and\
  \bibinfo {author} {\bibfnamefont {W.~B.}\ \bibnamefont {{Mori}}},\ }\bibfield
   {title} {\bibinfo {title} {{Ponderomotive force of quasiparticles in a
  plasma}},\ }\href {https://doi.org/10.1103/PhysRevE.59.2273} {\bibfield
  {journal} {\bibinfo  {journal} {\pre}\ }\textbf {\bibinfo {volume} {59}},\
  \bibinfo {pages} {2273} (\bibinfo {year} {1999}{\natexlab{a}})},\ \Eprint
  {https://arxiv.org/abs/physics/9807049} {arXiv:physics/9807049
  [physics.plasm-ph]} \BibitemShut {NoStop}%
\bibitem [{\citenamefont {{Silva}}\ \emph
  {et~al.}(1999{\natexlab{b}})\citenamefont {{Silva}}, \citenamefont
  {{Bingham}}, \citenamefont {{Dawson}}, \citenamefont {{Shukla}},
  \citenamefont {{Tsintsadze}},\ and\ \citenamefont
  {{Mendon{\c{c}}a}}}]{Silva1999D}%
  \BibitemOpen
  \bibfield  {author} {\bibinfo {author} {\bibfnamefont {L.~O.}\ \bibnamefont
  {{Silva}}}, \bibinfo {author} {\bibfnamefont {R.}~\bibnamefont {{Bingham}}},
  \bibinfo {author} {\bibfnamefont {J.~M.}\ \bibnamefont {{Dawson}}}, \bibinfo
  {author} {\bibfnamefont {P.~K.}\ \bibnamefont {{Shukla}}}, \bibinfo {author}
  {\bibfnamefont {N.~L.}\ \bibnamefont {{Tsintsadze}}},\ and\ \bibinfo {author}
  {\bibfnamefont {J.~T.}\ \bibnamefont {{Mendon{\c{c}}a}}},\ }\bibfield
  {title} {\bibinfo {title} {{Comment on ``Ponderomotive force due to
  neutrinos''}},\ }\href {https://doi.org/10.1103/PhysRevD.60.068701}
  {\bibfield  {journal} {\bibinfo  {journal} {\prd}\ }\textbf {\bibinfo
  {volume} {60}},\ \bibinfo {eid} {068701} (\bibinfo {year}
  {1999}{\natexlab{b}})},\ \Eprint {https://arxiv.org/abs/physics/9807050}
  {arXiv:physics/9807050 [physics.plasm-ph]} \BibitemShut {NoStop}%
\bibitem [{\citenamefont {{Silva}}\ \emph
  {et~al.}(1999{\natexlab{c}})\citenamefont {{Silva}}, \citenamefont
  {{Bingham}}, \citenamefont {{Dawson}}, \citenamefont {{Mendon{\c{c}}a}},\
  and\ \citenamefont {{Shukla}}}]{Silva1999L}%
  \BibitemOpen
  \bibfield  {author} {\bibinfo {author} {\bibfnamefont {L.~O.}\ \bibnamefont
  {{Silva}}}, \bibinfo {author} {\bibfnamefont {R.}~\bibnamefont {{Bingham}}},
  \bibinfo {author} {\bibfnamefont {J.~M.}\ \bibnamefont {{Dawson}}}, \bibinfo
  {author} {\bibfnamefont {J.~T.}\ \bibnamefont {{Mendon{\c{c}}a}}},\ and\
  \bibinfo {author} {\bibfnamefont {P.~K.}\ \bibnamefont {{Shukla}}},\
  }\bibfield  {title} {\bibinfo {title} {{Neutrino Driven Streaming
  Instabilities in a Dense Plasma}},\ }\href
  {https://doi.org/10.1103/PhysRevLett.83.2703} {\bibfield  {journal} {\bibinfo
   {journal} {\prl}\ }\textbf {\bibinfo {volume} {83}},\ \bibinfo {pages}
  {2703} (\bibinfo {year} {1999}{\natexlab{c}})}\BibitemShut {NoStop}%
\bibitem [{\citenamefont {{Silva}}\ \emph
  {et~al.}(2000{\natexlab{a}})\citenamefont {{Silva}}, \citenamefont
  {{Bingham}}, \citenamefont {{Dawson}}, \citenamefont {{Mendon{\c{c}}a}},\
  and\ \citenamefont {{Shukla}}}]{Silva2000}%
  \BibitemOpen
  \bibfield  {author} {\bibinfo {author} {\bibfnamefont {L.~O.}\ \bibnamefont
  {{Silva}}}, \bibinfo {author} {\bibfnamefont {R.}~\bibnamefont {{Bingham}}},
  \bibinfo {author} {\bibfnamefont {J.~M.}\ \bibnamefont {{Dawson}}}, \bibinfo
  {author} {\bibfnamefont {J.~T.}\ \bibnamefont {{Mendon{\c{c}}a}}},\ and\
  \bibinfo {author} {\bibfnamefont {P.~K.}\ \bibnamefont {{Shukla}}},\
  }\bibfield  {title} {\bibinfo {title} {{Collective neutrino-plasma
  interactions}},\ }\href {https://doi.org/10.1063/1.874037} {\bibfield
  {journal} {\bibinfo  {journal} {Physics of Plasmas}\ }\textbf {\bibinfo
  {volume} {7}},\ \bibinfo {pages} {2166} (\bibinfo {year}
  {2000}{\natexlab{a}})}\BibitemShut {NoStop}%
\bibitem [{\citenamefont {Bingham}\ \emph {et~al.}(2004)\citenamefont
  {Bingham}, \citenamefont {Silva}, \citenamefont {Mendon{\c{c}}a},
  \citenamefont {Shukla}, \citenamefont {Mori},\ and\ \citenamefont
  {Serbeto}}]{Bingham2004}%
  \BibitemOpen
  \bibfield  {author} {\bibinfo {author} {\bibfnamefont {R.}~\bibnamefont
  {Bingham}}, \bibinfo {author} {\bibfnamefont {L.~O.}\ \bibnamefont {Silva}},
  \bibinfo {author} {\bibfnamefont {J.~T.}\ \bibnamefont {Mendon{\c{c}}a}},
  \bibinfo {author} {\bibfnamefont {P.~K.}\ \bibnamefont {Shukla}}, \bibinfo
  {author} {\bibfnamefont {W.~B.}\ \bibnamefont {Mori}},\ and\ \bibinfo
  {author} {\bibfnamefont {A.}~\bibnamefont {Serbeto}},\ }\bibfield  {title}
  {\bibinfo {title} {Neutrino plasma coupling in dense astrophysical plasmas},\
  }\href {https://doi.org/10.1088/0741-3335/46/12b/028} {\bibfield  {journal}
  {\bibinfo  {journal} {Plasma Physics and Controlled Fusion}\ }\textbf
  {\bibinfo {volume} {46}},\ \bibinfo {pages} {B327} (\bibinfo {year}
  {2004})}\BibitemShut {NoStop}%
\bibitem [{\citenamefont {{Shukla}}\ \emph {et~al.}(1997)\citenamefont
  {{Shukla}}, \citenamefont {{Stenflo}}, \citenamefont {{Bingham}},
  \citenamefont {{Bethe}}, \citenamefont {{Dawson}},\ and\ \citenamefont
  {{Mendon{\c{c}}a}}}]{Shukla1997}%
  \BibitemOpen
  \bibfield  {author} {\bibinfo {author} {\bibfnamefont {P.~K.}\ \bibnamefont
  {{Shukla}}}, \bibinfo {author} {\bibfnamefont {L.}~\bibnamefont {{Stenflo}}},
  \bibinfo {author} {\bibfnamefont {R.}~\bibnamefont {{Bingham}}}, \bibinfo
  {author} {\bibfnamefont {H.~A.}\ \bibnamefont {{Bethe}}}, \bibinfo {author}
  {\bibfnamefont {J.~M.}\ \bibnamefont {{Dawson}}},\ and\ \bibinfo {author}
  {\bibfnamefont {J.~T.}\ \bibnamefont {{Mendon{\c{c}}a}}},\ }\bibfield
  {title} {\bibinfo {title} {{Generation of magnetic fields by nonuniform
  neutrino beams}},\ }\href {https://doi.org/10.1016/S0375-9601(97)00450-7}
  {\bibfield  {journal} {\bibinfo  {journal} {Physics Letters A}\ }\textbf
  {\bibinfo {volume} {233}},\ \bibinfo {pages} {181} (\bibinfo {year}
  {1997})}\BibitemShut {NoStop}%
\bibitem [{\citenamefont {Shukla}\ and\ \citenamefont
  {Stenflo}(1998)}]{Shukla1998}%
  \BibitemOpen
  \bibfield  {author} {\bibinfo {author} {\bibfnamefont {P.~K.}\ \bibnamefont
  {Shukla}}\ and\ \bibinfo {author} {\bibfnamefont {L.}~\bibnamefont
  {Stenflo}},\ }\bibfield  {title} {\bibinfo {title} {Intense magnetic fields
  produced by neutrino beams in supernovae},\ }\href
  {https://doi.org/10.1103/PhysRevE.57.2479} {\bibfield  {journal} {\bibinfo
  {journal} {Phys. Rev. E}\ }\textbf {\bibinfo {volume} {57}},\ \bibinfo
  {pages} {2479} (\bibinfo {year} {1998})}\BibitemShut {NoStop}%
\bibitem [{\citenamefont {Brizard}\ \emph {et~al.}(2000)\citenamefont
  {Brizard}, \citenamefont {Murayama},\ and\ \citenamefont
  {Wurtele}}]{Brizard2000}%
  \BibitemOpen
  \bibfield  {author} {\bibinfo {author} {\bibfnamefont {A.~J.}\ \bibnamefont
  {Brizard}}, \bibinfo {author} {\bibfnamefont {H.}~\bibnamefont {Murayama}},\
  and\ \bibinfo {author} {\bibfnamefont {J.~S.}\ \bibnamefont {Wurtele}},\
  }\bibfield  {title} {\bibinfo {title} {Magnetic field generation from
  self-consistent collective neutrino-plasma interactions},\ }\href
  {https://doi.org/10.1103/PhysRevE.61.4410} {\bibfield  {journal} {\bibinfo
  {journal} {Phys. Rev. E}\ }\textbf {\bibinfo {volume} {61}},\ \bibinfo
  {pages} {4410} (\bibinfo {year} {2000})}\BibitemShut {NoStop}%
\bibitem [{\citenamefont {{Turner}}\ and\ \citenamefont
  {{Widrow}}(1988)}]{Turner1988}%
  \BibitemOpen
  \bibfield  {author} {\bibinfo {author} {\bibfnamefont {M.~S.}\ \bibnamefont
  {{Turner}}}\ and\ \bibinfo {author} {\bibfnamefont {L.~M.}\ \bibnamefont
  {{Widrow}}},\ }\bibfield  {title} {\bibinfo {title} {{Inflation-produced,
  large-scale magnetic fields}},\ }\href
  {https://doi.org/10.1103/PhysRevD.37.2743} {\bibfield  {journal} {\bibinfo
  {journal} {\prd}\ }\textbf {\bibinfo {volume} {37}},\ \bibinfo {pages} {2743}
  (\bibinfo {year} {1988})}\BibitemShut {NoStop}%
\bibitem [{\citenamefont {Kronberg}(1994)}]{Kronberg1994}%
  \BibitemOpen
  \bibfield  {author} {\bibinfo {author} {\bibfnamefont {P.~P.}\ \bibnamefont
  {Kronberg}},\ }\bibfield  {title} {\bibinfo {title} {Extragalactic magnetic
  fields},\ }\href {https://doi.org/10.1088/0034-4885/57/4/001} {\bibfield
  {journal} {\bibinfo  {journal} {Reports on Progress in Physics}\ }\textbf
  {\bibinfo {volume} {57}},\ \bibinfo {pages} {325} (\bibinfo {year}
  {1994})}\BibitemShut {NoStop}%
\bibitem [{\citenamefont {{Enqvist}}(1998)}]{Enqvist1998}%
  \BibitemOpen
  \bibfield  {author} {\bibinfo {author} {\bibfnamefont {K.}~\bibnamefont
  {{Enqvist}}},\ }\bibfield  {title} {\bibinfo {title} {{Primordial Magnetic
  Fields}},\ }\href {https://doi.org/10.1142/S0218271898000243} {\bibfield
  {journal} {\bibinfo  {journal} {International Journal of Modern Physics D}\
  }\textbf {\bibinfo {volume} {7}},\ \bibinfo {pages} {331} (\bibinfo {year}
  {1998})},\ \Eprint {https://arxiv.org/abs/astro-ph/9803196}
  {arXiv:astro-ph/9803196 [astro-ph]} \BibitemShut {NoStop}%
\bibitem [{\citenamefont {{Son}}(1999)}]{Son1999}%
  \BibitemOpen
  \bibfield  {author} {\bibinfo {author} {\bibfnamefont {D.~T.}\ \bibnamefont
  {{Son}}},\ }\bibfield  {title} {\bibinfo {title} {{Magnetohydrodynamics of
  the early Universe and the evolution of primordial magnetic fields}},\ }\href
  {https://doi.org/10.1103/PhysRevD.59.063008} {\bibfield  {journal} {\bibinfo
  {journal} {\prd}\ }\textbf {\bibinfo {volume} {59}},\ \bibinfo {eid} {063008}
  (\bibinfo {year} {1999})},\ \Eprint {https://arxiv.org/abs/hep-ph/9803412}
  {arXiv:hep-ph/9803412 [hep-ph]} \BibitemShut {NoStop}%
\bibitem [{\citenamefont {{Giovannini}}(2004)}]{Giovannini2004}%
  \BibitemOpen
  \bibfield  {author} {\bibinfo {author} {\bibfnamefont {M.}~\bibnamefont
  {{Giovannini}}},\ }\bibfield  {title} {\bibinfo {title} {{The Magnetized
  Universe}},\ }\href {https://doi.org/10.1142/S0218271804004530} {\bibfield
  {journal} {\bibinfo  {journal} {International Journal of Modern Physics D}\
  }\textbf {\bibinfo {volume} {13}},\ \bibinfo {pages} {391} (\bibinfo {year}
  {2004})},\ \Eprint {https://arxiv.org/abs/astro-ph/0312614}
  {arXiv:astro-ph/0312614 [astro-ph]} \BibitemShut {NoStop}%
\bibitem [{\citenamefont {{Dolgov}}\ and\ \citenamefont
  {{Grasso}}(2001)}]{Dolgov2002}%
  \BibitemOpen
  \bibfield  {author} {\bibinfo {author} {\bibfnamefont {A.~D.}\ \bibnamefont
  {{Dolgov}}}\ and\ \bibinfo {author} {\bibfnamefont {D.}~\bibnamefont
  {{Grasso}}},\ }\bibfield  {title} {\bibinfo {title} {{Generation of Magnetic
  Fields and Gravitational Waves at Neutrino Decoupling}},\ }\href
  {https://doi.org/10.1103/PhysRevLett.88.011301} {\bibfield  {journal}
  {\bibinfo  {journal} {\prl}\ }\textbf {\bibinfo {volume} {88}},\ \bibinfo
  {eid} {011301} (\bibinfo {year} {2001})},\ \Eprint
  {https://arxiv.org/abs/astro-ph/0106154} {arXiv:astro-ph/0106154 [astro-ph]}
  \BibitemShut {NoStop}%
\bibitem [{\citenamefont {Semikoz}\ and\ \citenamefont
  {Sokoloff}(2004)}]{Semikoz2004}%
  \BibitemOpen
  \bibfield  {author} {\bibinfo {author} {\bibfnamefont {V.~B.}\ \bibnamefont
  {Semikoz}}\ and\ \bibinfo {author} {\bibfnamefont {D.~D.}\ \bibnamefont
  {Sokoloff}},\ }\bibfield  {title} {\bibinfo {title} {Large-scale magnetic
  field generation by $\ensuremath{\alpha}$ effect driven by collective
  neutrino-plasma interaction},\ }\href
  {https://doi.org/10.1103/PhysRevLett.92.131301} {\bibfield  {journal}
  {\bibinfo  {journal} {Phys. Rev. Lett.}\ }\textbf {\bibinfo {volume} {92}},\
  \bibinfo {pages} {131301} (\bibinfo {year} {2004})}\BibitemShut {NoStop}%
\bibitem [{\citenamefont {{Dvornikov}}\ and\ \citenamefont
  {{Semikoz}}(2014)}]{Dvornikov2014}%
  \BibitemOpen
  \bibfield  {author} {\bibinfo {author} {\bibfnamefont {M.}~\bibnamefont
  {{Dvornikov}}}\ and\ \bibinfo {author} {\bibfnamefont {V.~B.}\ \bibnamefont
  {{Semikoz}}},\ }\bibfield  {title} {\bibinfo {title} {{Instability of
  magnetic fields in electroweak plasma driven by neutrino asymmetries}},\
  }\href {https://doi.org/10.1088/1475-7516/2014/05/002} {\bibfield  {journal}
  {\bibinfo  {journal} {\jcap}\ }\textbf {\bibinfo {volume} {2014}},\ \bibinfo
  {eid} {002} (\bibinfo {year} {2014})},\ \Eprint
  {https://arxiv.org/abs/1311.5267} {arXiv:1311.5267 [hep-ph]} \BibitemShut
  {NoStop}%
\bibitem [{\citenamefont {Yamamoto}(2016)}]{Yamamoto2016}%
  \BibitemOpen
  \bibfield  {author} {\bibinfo {author} {\bibfnamefont {N.}~\bibnamefont
  {Yamamoto}},\ }\bibfield  {title} {\bibinfo {title} {Chiral transport of
  neutrinos in supernovae: Neutrino-induced fluid helicity and helical plasma
  instability},\ }\href {https://doi.org/10.1103/PhysRevD.93.065017} {\bibfield
   {journal} {\bibinfo  {journal} {Phys. Rev. D}\ }\textbf {\bibinfo {volume}
  {93}},\ \bibinfo {pages} {065017} (\bibinfo {year} {2016})}\BibitemShut
  {NoStop}%
\bibitem [{\citenamefont {{Pandey}}\ \emph {et~al.}(2020)\citenamefont
  {{Pandey}}, \citenamefont {{Natwariya}},\ and\ \citenamefont
  {{Bhatt}}}]{Pandey2020}%
  \BibitemOpen
  \bibfield  {author} {\bibinfo {author} {\bibfnamefont {A.~K.}\ \bibnamefont
  {{Pandey}}}, \bibinfo {author} {\bibfnamefont {P.~K.}\ \bibnamefont
  {{Natwariya}}},\ and\ \bibinfo {author} {\bibfnamefont {J.~R.}\ \bibnamefont
  {{Bhatt}}},\ }\bibfield  {title} {\bibinfo {title} {{Magnetic fields in a hot
  dense neutrino plasma and the gravitational waves}},\ }\href
  {https://doi.org/10.1103/PhysRevD.101.023531} {\bibfield  {journal} {\bibinfo
   {journal} {\prd}\ }\textbf {\bibinfo {volume} {101}},\ \bibinfo {eid}
  {023531} (\bibinfo {year} {2020})},\ \Eprint
  {https://arxiv.org/abs/1911.05412} {arXiv:1911.05412 [astro-ph.CO]}
  \BibitemShut {NoStop}%
\bibitem [{\citenamefont {{Semikoz}}(1987)}]{Semikoz1987}%
  \BibitemOpen
  \bibfield  {author} {\bibinfo {author} {\bibfnamefont {V.~B.}\ \bibnamefont
  {{Semikoz}}},\ }\bibfield  {title} {\bibinfo {title} {{Kinetics of a lepton
  plasma in the standard model of electroweak interactions}},\ }\href
  {https://doi.org/10.1016/0378-4371(87)90022-7} {\bibfield  {journal}
  {\bibinfo  {journal} {Physica A Statistical Mechanics and its Applications}\
  }\textbf {\bibinfo {volume} {142}},\ \bibinfo {pages} {157} (\bibinfo {year}
  {1987})}\BibitemShut {NoStop}%
\bibitem [{\citenamefont {{Silva}}\ \emph
  {et~al.}(2000{\natexlab{b}})\citenamefont {{Silva}}, \citenamefont
  {{Bingham}}, \citenamefont {{Dawson}}, \citenamefont {{Mendon{\c{c}}a}},\
  and\ \citenamefont {{Shukla}}}]{Silva2000b}%
  \BibitemOpen
  \bibfield  {author} {\bibinfo {author} {\bibfnamefont {L.~O.}\ \bibnamefont
  {{Silva}}}, \bibinfo {author} {\bibfnamefont {R.}~\bibnamefont {{Bingham}}},
  \bibinfo {author} {\bibfnamefont {J.~M.}\ \bibnamefont {{Dawson}}}, \bibinfo
  {author} {\bibfnamefont {J.~T.}\ \bibnamefont {{Mendon{\c{c}}a}}},\ and\
  \bibinfo {author} {\bibfnamefont {P.~K.}\ \bibnamefont {{Shukla}}},\
  }\bibfield  {title} {\bibinfo {title} {{Neutrino Kinetics in Dense
  Astrophysical Plasmas}},\ }\href {https://doi.org/10.1086/313335} {\bibfield
  {journal} {\bibinfo  {journal} {\apjs}\ }\textbf {\bibinfo {volume} {127}},\
  \bibinfo {pages} {481} (\bibinfo {year} {2000}{\natexlab{b}})}\BibitemShut
  {NoStop}%
\bibitem [{\citenamefont {Brizard}\ and\ \citenamefont
  {Wurtele}(1999)}]{Brizard1999}%
  \BibitemOpen
  \bibfield  {author} {\bibinfo {author} {\bibfnamefont {A.~J.}\ \bibnamefont
  {Brizard}}\ and\ \bibinfo {author} {\bibfnamefont {J.~S.}\ \bibnamefont
  {Wurtele}},\ }\bibfield  {title} {\bibinfo {title} {Lagrangian formulation
  for neutrino–plasma interactions},\ }\href
  {https://doi.org/10.1063/1.873373} {\bibfield  {journal} {\bibinfo  {journal}
  {Physics of Plasmas}\ }\textbf {\bibinfo {volume} {6}},\ \bibinfo {pages}
  {1323} (\bibinfo {year} {1999})},\ \Eprint
  {https://arxiv.org/abs/https://doi.org/10.1063/1.873373}
  {https://doi.org/10.1063/1.873373} \BibitemShut {NoStop}%
\bibitem [{\citenamefont {Shukla}(2003)}]{Shukla2003}%
  \BibitemOpen
  \bibfield  {author} {\bibinfo {author} {\bibfnamefont {P.}~\bibnamefont
  {Shukla}},\ }\bibfield  {title} {\bibinfo {title} {Generation of magnetic
  fields in the early universe},\ }\href
  {https://doi.org/https://doi.org/10.1016/S0375-9601(03)00336-0} {\bibfield
  {journal} {\bibinfo  {journal} {Physics Letters A}\ }\textbf {\bibinfo
  {volume} {310}},\ \bibinfo {pages} {182} (\bibinfo {year}
  {2003})}\BibitemShut {NoStop}%
\bibitem [{\citenamefont {{Haas}}\ \emph {et~al.}(2016)\citenamefont {{Haas}},
  \citenamefont {{Pascoal}},\ and\ \citenamefont
  {{Mendon{\c{c}}a}}}]{Haas2016}%
  \BibitemOpen
  \bibfield  {author} {\bibinfo {author} {\bibfnamefont {F.}~\bibnamefont
  {{Haas}}}, \bibinfo {author} {\bibfnamefont {K.~A.}\ \bibnamefont
  {{Pascoal}}},\ and\ \bibinfo {author} {\bibfnamefont {J.~T.}\ \bibnamefont
  {{Mendon{\c{c}}a}}},\ }\bibfield  {title} {\bibinfo {title} {{Neutrino
  magnetohydrodynamics}},\ }\href {https://doi.org/10.1063/1.4939535}
  {\bibfield  {journal} {\bibinfo  {journal} {Physics of Plasmas}\ }\textbf
  {\bibinfo {volume} {23}},\ \bibinfo {eid} {012104} (\bibinfo {year}
  {2016})}\BibitemShut {NoStop}%
\bibitem [{\citenamefont {Wolfenstein}(1978)}]{Wolfenstein1978}%
  \BibitemOpen
  \bibfield  {author} {\bibinfo {author} {\bibfnamefont {L.}~\bibnamefont
  {Wolfenstein}},\ }\bibfield  {title} {\bibinfo {title} {Neutrino oscillations
  in matter},\ }\href {https://doi.org/10.1103/PhysRevD.17.2369} {\bibfield
  {journal} {\bibinfo  {journal} {Phys. Rev. D}\ }\textbf {\bibinfo {volume}
  {17}},\ \bibinfo {pages} {2369} (\bibinfo {year} {1978})}\BibitemShut
  {NoStop}%
\bibitem [{\citenamefont {{Tajima}}\ and\ \citenamefont
  {{Shibata}}(2002)}]{Tajima2002}%
  \BibitemOpen
  \bibfield  {author} {\bibinfo {author} {\bibfnamefont {T.}~\bibnamefont
  {{Tajima}}}\ and\ \bibinfo {author} {\bibfnamefont {K.}~\bibnamefont
  {{Shibata}}},\ }\href@noop {} {\emph {\bibinfo {title} {{Plasma
  astrophysics}}}}\ (\bibinfo {year} {2002})\BibitemShut {NoStop}%
\bibitem [{Note1()}]{Note1}%
  \BibitemOpen
  \bibinfo {note} {Note that we use the definition of $\protect \mathcal {L}$
  as the time derivative of the action $S$ with respect to coordinate time and
  not proper time.}\BibitemShut {Stop}%
\bibitem [{\citenamefont {{Tajima}}\ and\ \citenamefont
  {{Dawson}}(1979)}]{Tajima1979}%
  \BibitemOpen
  \bibfield  {author} {\bibinfo {author} {\bibfnamefont {T.}~\bibnamefont
  {{Tajima}}}\ and\ \bibinfo {author} {\bibfnamefont {J.~M.}\ \bibnamefont
  {{Dawson}}},\ }\bibfield  {title} {\bibinfo {title} {{Laser electron
  accelerator}},\ }\href {https://doi.org/10.1103/PhysRevLett.43.267}
  {\bibfield  {journal} {\bibinfo  {journal} {\prl}\ }\textbf {\bibinfo
  {volume} {43}},\ \bibinfo {pages} {267} (\bibinfo {year} {1979})}\BibitemShut
  {NoStop}%
\bibitem [{\citenamefont {{Esarey}}\ \emph {et~al.}(2009)\citenamefont
  {{Esarey}}, \citenamefont {{Schroeder}},\ and\ \citenamefont
  {{Leemans}}}]{Esarey2009}%
  \BibitemOpen
  \bibfield  {author} {\bibinfo {author} {\bibfnamefont {E.}~\bibnamefont
  {{Esarey}}}, \bibinfo {author} {\bibfnamefont {C.~B.}\ \bibnamefont
  {{Schroeder}}},\ and\ \bibinfo {author} {\bibfnamefont {W.~P.}\ \bibnamefont
  {{Leemans}}},\ }\bibfield  {title} {\bibinfo {title} {{Physics of
  laser-driven plasma-based electron accelerators}},\ }\href
  {https://doi.org/10.1103/RevModPhys.81.1229} {\bibfield  {journal} {\bibinfo
  {journal} {Reviews of Modern Physics}\ }\textbf {\bibinfo {volume} {81}},\
  \bibinfo {pages} {1229} (\bibinfo {year} {2009})}\BibitemShut {NoStop}%
\bibitem [{\citenamefont {Gibbon}(2005)}]{Gibbon2005}%
  \BibitemOpen
  \bibfield  {author} {\bibinfo {author} {\bibfnamefont {P.}~\bibnamefont
  {Gibbon}},\ }\href {https://doi.org/10.1142/p116} {\emph {\bibinfo {title}
  {Short Pulse Laser Interactions with Matter}}}\ (\bibinfo  {publisher}
  {Published by Imperial College Press and distributed by World Scientific
  Publishing Co.},\ \bibinfo {year} {2005})\ \Eprint
  {https://arxiv.org/abs/https://www.worldscientific.com/doi/pdf/10.1142/p116}
  {https://www.worldscientific.com/doi/pdf/10.1142/p116} \BibitemShut {NoStop}%
\bibitem [{\citenamefont {{Mulser}}\ and\ \citenamefont
  {{Bauer}}(2010)}]{Mulser2010}%
  \BibitemOpen
  \bibfield  {author} {\bibinfo {author} {\bibfnamefont {P.}~\bibnamefont
  {{Mulser}}}\ and\ \bibinfo {author} {\bibfnamefont {D.}~\bibnamefont
  {{Bauer}}},\ }\href {https://doi.org/10.1007/978-3-540-46065-7} {\emph
  {\bibinfo {title} {{High Power Laser-Matter Interaction}}}},\ Vol.\ \bibinfo
  {volume} {238}\ (\bibinfo {year} {2010})\BibitemShut {NoStop}%
\bibitem [{\citenamefont {{Thorne}}\ and\ \citenamefont
  {{MacDonald}}(1982)}]{Thorne1982}%
  \BibitemOpen
  \bibfield  {author} {\bibinfo {author} {\bibfnamefont {K.~S.}\ \bibnamefont
  {{Thorne}}}\ and\ \bibinfo {author} {\bibfnamefont {D.}~\bibnamefont
  {{MacDonald}}},\ }\bibfield  {title} {\bibinfo {title} {{Electrodynamics in
  Curved Spacetime - 3+1 Formulation}},\ }\href
  {https://doi.org/10.1093/mnras/198.2.339} {\bibfield  {journal} {\bibinfo
  {journal} {\mnras}\ }\textbf {\bibinfo {volume} {198}},\ \bibinfo {pages}
  {339} (\bibinfo {year} {1982})}\BibitemShut {NoStop}%
\bibitem [{\citenamefont {{Holcomb}}\ and\ \citenamefont
  {{Tajima}}(1989)}]{Holcomb1989}%
  \BibitemOpen
  \bibfield  {author} {\bibinfo {author} {\bibfnamefont {K.~A.}\ \bibnamefont
  {{Holcomb}}}\ and\ \bibinfo {author} {\bibfnamefont {T.}~\bibnamefont
  {{Tajima}}},\ }\bibfield  {title} {\bibinfo {title} {{General-relativistic
  plasma physics in the early Universe}},\ }\href
  {https://doi.org/10.1103/PhysRevD.40.3809} {\bibfield  {journal} {\bibinfo
  {journal} {\prd}\ }\textbf {\bibinfo {volume} {40}},\ \bibinfo {pages} {3809}
  (\bibinfo {year} {1989})}\BibitemShut {NoStop}%
\bibitem [{\citenamefont {Lifshitz}\ and\ \citenamefont
  {Pitaevskii}(1981)}]{Lifshitz:99987}%
  \BibitemOpen
  \bibfield  {author} {\bibinfo {author} {\bibfnamefont {E.~M.}\ \bibnamefont
  {Lifshitz}}\ and\ \bibinfo {author} {\bibfnamefont {L.~P.}\ \bibnamefont
  {Pitaevskii}},\ }\href {https://cds.cern.ch/record/99987} {\emph {\bibinfo
  {title} {{Physical kinetics}}}},\ Course of theoretical physics\ (\bibinfo
  {publisher} {Pergamon},\ \bibinfo {address} {Oxford},\ \bibinfo {year}
  {1981})\ \bibinfo {note} {translated from the Russian by J B Sykes and R N
  Franklin}\BibitemShut {NoStop}%
\bibitem [{\citenamefont {{Ehlers}}(1971)}]{Ehlers1971}%
  \BibitemOpen
  \bibfield  {author} {\bibinfo {author} {\bibfnamefont {J.}~\bibnamefont
  {{Ehlers}}},\ }\bibfield  {title} {\bibinfo {title} {{General relativity and
  kinetic theory.}},\ }in\ \href@noop {} {\emph {\bibinfo {booktitle} {General
  Relativity and Cosmology}}}\ (\bibinfo {year} {1971})\ pp.\ \bibinfo {pages}
  {1--70}\BibitemShut {NoStop}%
\bibitem [{\citenamefont {{Dettmann}}\ \emph {et~al.}(1993)\citenamefont
  {{Dettmann}}, \citenamefont {{Frankel}},\ and\ \citenamefont
  {{Kowalenko}}}]{Dettmann1993}%
  \BibitemOpen
  \bibfield  {author} {\bibinfo {author} {\bibfnamefont {C.~P.}\ \bibnamefont
  {{Dettmann}}}, \bibinfo {author} {\bibfnamefont {N.~E.}\ \bibnamefont
  {{Frankel}}},\ and\ \bibinfo {author} {\bibfnamefont {V.}~\bibnamefont
  {{Kowalenko}}},\ }\bibfield  {title} {\bibinfo {title} {{Plasma
  electrodynamics in the expanding Universe}},\ }\href
  {https://doi.org/10.1103/PhysRevD.48.5655} {\bibfield  {journal} {\bibinfo
  {journal} {\prd}\ }\textbf {\bibinfo {volume} {48}},\ \bibinfo {pages} {5655}
  (\bibinfo {year} {1993})}\BibitemShut {NoStop}%
\bibitem [{\citenamefont {De~Groot}(1980)}]{DeGroot1980}%
  \BibitemOpen
  \bibfield  {author} {\bibinfo {author} {\bibfnamefont {S.~R.}\ \bibnamefont
  {De~Groot}},\ }\href@noop {} {\emph {\bibinfo {title} {{Relativistic Kinetic
  Theory. Principles and Applications}}}},\ edited by\ \bibinfo {editor}
  {\bibfnamefont {W.~A.}\ \bibnamefont {Van~Leeuwen}}\ and\ \bibinfo {editor}
  {\bibfnamefont {C.~G.}\ \bibnamefont {Van~Weert}}\ (\bibinfo {year}
  {1980})\BibitemShut {NoStop}%
\bibitem [{\citenamefont {{Kolb}}\ and\ \citenamefont
  {{Turner}}(1990)}]{Kolb1990}%
  \BibitemOpen
  \bibfield  {author} {\bibinfo {author} {\bibfnamefont {E.~W.}\ \bibnamefont
  {{Kolb}}}\ and\ \bibinfo {author} {\bibfnamefont {M.~S.}\ \bibnamefont
  {{Turner}}},\ }\href@noop {} {\emph {\bibinfo {title} {{The early
  universe}}}},\ Vol.~\bibinfo {volume} {69}\ (\bibinfo {year}
  {1990})\BibitemShut {NoStop}%
\bibitem [{\citenamefont {{Gailis}}\ \emph {et~al.}(1995)\citenamefont
  {{Gailis}}, \citenamefont {{Frankel}},\ and\ \citenamefont
  {{Dettmann}}}]{Gailis1995}%
  \BibitemOpen
  \bibfield  {author} {\bibinfo {author} {\bibfnamefont {R.~M.}\ \bibnamefont
  {{Gailis}}}, \bibinfo {author} {\bibfnamefont {N.~E.}\ \bibnamefont
  {{Frankel}}},\ and\ \bibinfo {author} {\bibfnamefont {C.~P.}\ \bibnamefont
  {{Dettmann}}},\ }\bibfield  {title} {\bibinfo {title} {{Magnetohydrodynamics
  in the expanding Universe}},\ }\href
  {https://doi.org/10.1103/PhysRevD.52.6901} {\bibfield  {journal} {\bibinfo
  {journal} {\prd}\ }\textbf {\bibinfo {volume} {52}},\ \bibinfo {pages} {6901}
  (\bibinfo {year} {1995})}\BibitemShut {NoStop}%
\bibitem [{\citenamefont {{Brandenburg}}\ \emph {et~al.}(1996)\citenamefont
  {{Brandenburg}}, \citenamefont {{Enqvist}},\ and\ \citenamefont
  {{Olesen}}}]{Brandenburg1996}%
  \BibitemOpen
  \bibfield  {author} {\bibinfo {author} {\bibfnamefont {A.}~\bibnamefont
  {{Brandenburg}}}, \bibinfo {author} {\bibfnamefont {K.}~\bibnamefont
  {{Enqvist}}},\ and\ \bibinfo {author} {\bibfnamefont {P.}~\bibnamefont
  {{Olesen}}},\ }\bibfield  {title} {\bibinfo {title} {{Large-scale magnetic
  fields from hydromagnetic turbulence in the very early universe}},\ }\href
  {https://doi.org/10.1103/PhysRevD.54.1291} {\bibfield  {journal} {\bibinfo
  {journal} {\prd}\ }\textbf {\bibinfo {volume} {54}},\ \bibinfo {pages} {1291}
  (\bibinfo {year} {1996})},\ \Eprint {https://arxiv.org/abs/astro-ph/9602031}
  {arXiv:astro-ph/9602031 [astro-ph]} \BibitemShut {NoStop}%
\bibitem [{\citenamefont {Lighthill}(1960)}]{Lighthill1960}%
  \BibitemOpen
  \bibfield  {author} {\bibinfo {author} {\bibfnamefont {M.~J.}\ \bibnamefont
  {Lighthill}},\ }\bibfield  {title} {\bibinfo {title} {Studies on
  magneto-hydrodynamic waves and other anisotropic wave motions},\ }\href
  {http://www.jstor.org/stable/73169} {\bibfield  {journal} {\bibinfo
  {journal} {Philosophical Transactions of the Royal Society of London. Series
  A, Mathematical and Physical Sciences}\ }\textbf {\bibinfo {volume} {252}},\
  \bibinfo {pages} {397} (\bibinfo {year} {1960})}\BibitemShut {NoStop}%
\bibitem [{\citenamefont {{Kingsep}}\ \emph {et~al.}(1987)\citenamefont
  {{Kingsep}}, \citenamefont {{Chukbar}},\ and\ \citenamefont
  {{Ian'kov}}}]{Kingsep1987}%
  \BibitemOpen
  \bibfield  {author} {\bibinfo {author} {\bibfnamefont {A.~S.}\ \bibnamefont
  {{Kingsep}}}, \bibinfo {author} {\bibfnamefont {K.~V.}\ \bibnamefont
  {{Chukbar}}},\ and\ \bibinfo {author} {\bibfnamefont {V.~V.}\ \bibnamefont
  {{Ian'kov}}},\ }\bibfield  {title} {\bibinfo {title} {{Electron
  magnetohydrodynamics}},\ }\href@noop {} {\bibfield  {journal} {\bibinfo
  {journal} {Voprosy Teorii Plazmy}\ }\textbf {\bibinfo {volume} {16}},\
  \bibinfo {pages} {209} (\bibinfo {year} {1987})}\BibitemShut {NoStop}%
\bibitem [{\citenamefont {{Gordeev}}\ \emph {et~al.}(1994)\citenamefont
  {{Gordeev}}, \citenamefont {{Kingsep}},\ and\ \citenamefont
  {{Rudakov}}}]{Gordeev1994}%
  \BibitemOpen
  \bibfield  {author} {\bibinfo {author} {\bibfnamefont {A.~V.}\ \bibnamefont
  {{Gordeev}}}, \bibinfo {author} {\bibfnamefont {A.~S.}\ \bibnamefont
  {{Kingsep}}},\ and\ \bibinfo {author} {\bibfnamefont {L.~I.}\ \bibnamefont
  {{Rudakov}}},\ }\bibfield  {title} {\bibinfo {title} {{Electron
  magnetohydrodynamics}},\ }\href
  {https://doi.org/10.1016/0370-1573(94)90097-3} {\bibfield  {journal}
  {\bibinfo  {journal} {Physics Reports}\ }\textbf {\bibinfo {volume} {243}},\
  \bibinfo {pages} {215} (\bibinfo {year} {1994})}\BibitemShut {NoStop}%
\bibitem [{\citenamefont {Biskamp}\ \emph {et~al.}(1999)\citenamefont
  {Biskamp}, \citenamefont {Schwarz}, \citenamefont {Zeiler}, \citenamefont
  {Celani},\ and\ \citenamefont {Drake}}]{Biskampd1999}%
  \BibitemOpen
  \bibfield  {author} {\bibinfo {author} {\bibfnamefont {D.}~\bibnamefont
  {Biskamp}}, \bibinfo {author} {\bibfnamefont {E.}~\bibnamefont {Schwarz}},
  \bibinfo {author} {\bibfnamefont {A.}~\bibnamefont {Zeiler}}, \bibinfo
  {author} {\bibfnamefont {A.}~\bibnamefont {Celani}},\ and\ \bibinfo {author}
  {\bibfnamefont {J.~F.}\ \bibnamefont {Drake}},\ }\bibfield  {title} {\bibinfo
  {title} {Electron magnetohydrodynamic turbulence},\ }\href
  {https://doi.org/10.1063/1.873312} {\bibfield  {journal} {\bibinfo  {journal}
  {Physics of Plasmas}\ }\textbf {\bibinfo {volume} {6}},\ \bibinfo {pages}
  {751} (\bibinfo {year} {1999})},\ \Eprint
  {https://arxiv.org/abs/https://doi.org/10.1063/1.873312}
  {https://doi.org/10.1063/1.873312} \BibitemShut {NoStop}%
\bibitem [{\citenamefont {{Sigl}}\ \emph {et~al.}(1997)\citenamefont {{Sigl}},
  \citenamefont {{Olinto}},\ and\ \citenamefont {{Jedamzik}}}]{Sigl1997}%
  \BibitemOpen
  \bibfield  {author} {\bibinfo {author} {\bibfnamefont {G.}~\bibnamefont
  {{Sigl}}}, \bibinfo {author} {\bibfnamefont {A.~V.}\ \bibnamefont
  {{Olinto}}},\ and\ \bibinfo {author} {\bibfnamefont {K.}~\bibnamefont
  {{Jedamzik}}},\ }\bibfield  {title} {\bibinfo {title} {{Primordial magnetic
  fields from cosmological first order phase transitions}},\ }\href
  {https://doi.org/10.1103/PhysRevD.55.4582} {\bibfield  {journal} {\bibinfo
  {journal} {\prd}\ }\textbf {\bibinfo {volume} {55}},\ \bibinfo {pages} {4582}
  (\bibinfo {year} {1997})},\ \Eprint {https://arxiv.org/abs/astro-ph/9610201}
  {arXiv:astro-ph/9610201 [astro-ph]} \BibitemShut {NoStop}%
\bibitem [{\citenamefont {{Biermann}}(1950)}]{Biermann1950}%
  \BibitemOpen
  \bibfield  {author} {\bibinfo {author} {\bibfnamefont {L.}~\bibnamefont
  {{Biermann}}},\ }\bibfield  {title} {\bibinfo {title} {{{\"U}ber den Ursprung
  der Magnetfelder auf Sternen und im interstellaren Raum (miteinem Anhang von
  A. Schl{\"u}ter)}},\ }\href@noop {} {\bibfield  {journal} {\bibinfo
  {journal} {Zeitschrift Naturforschung Teil A}\ }\textbf {\bibinfo {volume}
  {5}},\ \bibinfo {pages} {65} (\bibinfo {year} {1950})}\BibitemShut {NoStop}%
\bibitem [{\citenamefont {{Harrison}}(1970)}]{Harrison1970}%
  \BibitemOpen
  \bibfield  {author} {\bibinfo {author} {\bibfnamefont {E.~R.}\ \bibnamefont
  {{Harrison}}},\ }\bibfield  {title} {\bibinfo {title} {{Generation of
  magnetic fields in the radiation ERA}},\ }\href
  {https://doi.org/10.1093/mnras/147.3.279} {\bibfield  {journal} {\bibinfo
  {journal} {\mnras}\ }\textbf {\bibinfo {volume} {147}},\ \bibinfo {pages}
  {279} (\bibinfo {year} {1970})}\BibitemShut {NoStop}%
\bibitem [{\citenamefont {{Grasso}}\ and\ \citenamefont
  {{Rubinstein}}(2001)}]{Grasso2001}%
  \BibitemOpen
  \bibfield  {author} {\bibinfo {author} {\bibfnamefont {D.}~\bibnamefont
  {{Grasso}}}\ and\ \bibinfo {author} {\bibfnamefont {H.~R.}\ \bibnamefont
  {{Rubinstein}}},\ }\bibfield  {title} {\bibinfo {title} {{Magnetic fields in
  the early Universe}},\ }\href {https://doi.org/10.1016/S0370-1573(00)00110-1}
  {\bibfield  {journal} {\bibinfo  {journal} {\physrep}\ }\textbf {\bibinfo
  {volume} {348}},\ \bibinfo {pages} {163} (\bibinfo {year} {2001})},\ \Eprint
  {https://arxiv.org/abs/astro-ph/0009061} {arXiv:astro-ph/0009061 [astro-ph]}
  \BibitemShut {NoStop}%
\bibitem [{\citenamefont {{Durrer}}\ and\ \citenamefont
  {{Neronov}}(2013)}]{Durrer2013}%
  \BibitemOpen
  \bibfield  {author} {\bibinfo {author} {\bibfnamefont {R.}~\bibnamefont
  {{Durrer}}}\ and\ \bibinfo {author} {\bibfnamefont {A.}~\bibnamefont
  {{Neronov}}},\ }\bibfield  {title} {\bibinfo {title} {{Cosmological magnetic
  fields: their generation, evolution and observation}},\ }\href
  {https://doi.org/10.1007/s00159-013-0062-7} {\bibfield  {journal} {\bibinfo
  {journal} {\aapr}\ }\textbf {\bibinfo {volume} {21}},\ \bibinfo {eid} {62}
  (\bibinfo {year} {2013})},\ \Eprint {https://arxiv.org/abs/1303.7121}
  {arXiv:1303.7121 [astro-ph.CO]} \BibitemShut {NoStop}%
\bibitem [{\citenamefont {Langacker}\ \emph {et~al.}(1983)\citenamefont
  {Langacker}, \citenamefont {Segrè},\ and\ \citenamefont
  {Soni}}]{Langacker1983}%
  \BibitemOpen
  \bibfield  {author} {\bibinfo {author} {\bibfnamefont {P.}~\bibnamefont
  {Langacker}}, \bibinfo {author} {\bibfnamefont {G.}~\bibnamefont {Segrè}},\
  and\ \bibinfo {author} {\bibfnamefont {S.}~\bibnamefont {Soni}},\ }\bibfield
  {title} {\bibinfo {title} {The lepton asymmetry of the universe},\ }\href
  {https://doi.org/10.1063/1.34026} {\bibfield  {journal} {\bibinfo  {journal}
  {AIP Conference Proceedings}\ }\textbf {\bibinfo {volume} {99}},\ \bibinfo
  {pages} {76} (\bibinfo {year} {1983})},\ \Eprint
  {https://arxiv.org/abs/https://aip.scitation.org/doi/pdf/10.1063/1.34026}
  {https://aip.scitation.org/doi/pdf/10.1063/1.34026} \BibitemShut {NoStop}%
\bibitem [{\citenamefont {{Aoki}}\ \emph {et~al.}(2006)\citenamefont {{Aoki}},
  \citenamefont {{Endr{\H{o}}di}}, \citenamefont {{Fodor}}, \citenamefont
  {{Katz}},\ and\ \citenamefont {{Szab{\'o}}}}]{Aoki2006}%
  \BibitemOpen
  \bibfield  {author} {\bibinfo {author} {\bibfnamefont {Y.}~\bibnamefont
  {{Aoki}}}, \bibinfo {author} {\bibfnamefont {G.}~\bibnamefont
  {{Endr{\H{o}}di}}}, \bibinfo {author} {\bibfnamefont {Z.}~\bibnamefont
  {{Fodor}}}, \bibinfo {author} {\bibfnamefont {S.~D.}\ \bibnamefont
  {{Katz}}},\ and\ \bibinfo {author} {\bibfnamefont {K.~K.}\ \bibnamefont
  {{Szab{\'o}}}},\ }\bibfield  {title} {\bibinfo {title} {{The order of the
  quantum chromodynamics transition predicted by the standard model of particle
  physics}},\ }\href {https://doi.org/10.1038/nature05120} {\bibfield
  {journal} {\bibinfo  {journal} {\nat}\ }\textbf {\bibinfo {volume} {443}},\
  \bibinfo {pages} {675} (\bibinfo {year} {2006})},\ \Eprint
  {https://arxiv.org/abs/hep-lat/0611014} {arXiv:hep-lat/0611014 [hep-lat]}
  \BibitemShut {NoStop}%
\bibitem [{\citenamefont {Miniati}\ \emph {et~al.}(2018)\citenamefont
  {Miniati}, \citenamefont {Gregori}, \citenamefont {Reville},\ and\
  \citenamefont {Sarkar}}]{Miniati2018}%
  \BibitemOpen
  \bibfield  {author} {\bibinfo {author} {\bibfnamefont {F.}~\bibnamefont
  {Miniati}}, \bibinfo {author} {\bibfnamefont {G.}~\bibnamefont {Gregori}},
  \bibinfo {author} {\bibfnamefont {B.}~\bibnamefont {Reville}},\ and\ \bibinfo
  {author} {\bibfnamefont {S.}~\bibnamefont {Sarkar}},\ }\bibfield  {title}
  {\bibinfo {title} {Axion-driven cosmic magnetogenesis during the qcd
  crossover},\ }\href {https://doi.org/10.1103/PhysRevLett.121.021301}
  {\bibfield  {journal} {\bibinfo  {journal} {Phys. Rev. Lett.}\ }\textbf
  {\bibinfo {volume} {121}},\ \bibinfo {pages} {021301} (\bibinfo {year}
  {2018})}\BibitemShut {NoStop}%
\bibitem [{\citenamefont {Ejiri}\ \emph {et~al.}(2006)\citenamefont {Ejiri},
  \citenamefont {Karsch},\ and\ \citenamefont {Redlich}}]{Ejiri2006}%
  \BibitemOpen
  \bibfield  {author} {\bibinfo {author} {\bibfnamefont {S.}~\bibnamefont
  {Ejiri}}, \bibinfo {author} {\bibfnamefont {F.}~\bibnamefont {Karsch}},\ and\
  \bibinfo {author} {\bibfnamefont {K.}~\bibnamefont {Redlich}},\ }\bibfield
  {title} {\bibinfo {title} {Hadronic fluctuations at the qcd phase
  transition},\ }\href
  {https://doi.org/https://doi.org/10.1016/j.physletb.2005.11.083} {\bibfield
  {journal} {\bibinfo  {journal} {Physics Letters B}\ }\textbf {\bibinfo
  {volume} {633}},\ \bibinfo {pages} {275} (\bibinfo {year}
  {2006})}\BibitemShut {NoStop}%
\bibitem [{\citenamefont {Bazavov}\ \emph {et~al.}(2013)\citenamefont
  {Bazavov}, \citenamefont {Ding}, \citenamefont {Hegde}, \citenamefont
  {Karsch}, \citenamefont {Miao}, \citenamefont {Mukherjee}, \citenamefont
  {Petreczky}, \citenamefont {Schmidt},\ and\ \citenamefont
  {Velytsky}}]{Bazavov2013}%
  \BibitemOpen
  \bibfield  {author} {\bibinfo {author} {\bibfnamefont {A.}~\bibnamefont
  {Bazavov}}, \bibinfo {author} {\bibfnamefont {H.-T.}\ \bibnamefont {Ding}},
  \bibinfo {author} {\bibfnamefont {P.}~\bibnamefont {Hegde}}, \bibinfo
  {author} {\bibfnamefont {F.}~\bibnamefont {Karsch}}, \bibinfo {author}
  {\bibfnamefont {C.}~\bibnamefont {Miao}}, \bibinfo {author} {\bibfnamefont
  {S.}~\bibnamefont {Mukherjee}}, \bibinfo {author} {\bibfnamefont
  {P.}~\bibnamefont {Petreczky}}, \bibinfo {author} {\bibfnamefont
  {C.}~\bibnamefont {Schmidt}},\ and\ \bibinfo {author} {\bibfnamefont
  {A.}~\bibnamefont {Velytsky}},\ }\bibfield  {title} {\bibinfo {title} {Quark
  number susceptibilities at high temperatures},\ }\href
  {https://doi.org/10.1103/PhysRevD.88.094021} {\bibfield  {journal} {\bibinfo
  {journal} {Phys. Rev. D}\ }\textbf {\bibinfo {volume} {88}},\ \bibinfo
  {pages} {094021} (\bibinfo {year} {2013})}\BibitemShut {NoStop}%
\bibitem [{\citenamefont {Bellwied}\ \emph {et~al.}(2015)\citenamefont
  {Bellwied}, \citenamefont {Bors\'anyi}, \citenamefont {Fodor}, \citenamefont
  {Katz}, \citenamefont {P\'asztor}, \citenamefont {Ratti},\ and\ \citenamefont
  {Szab\'o}}]{Bellwied2015}%
  \BibitemOpen
  \bibfield  {author} {\bibinfo {author} {\bibfnamefont {R.}~\bibnamefont
  {Bellwied}}, \bibinfo {author} {\bibfnamefont {S.}~\bibnamefont
  {Bors\'anyi}}, \bibinfo {author} {\bibfnamefont {Z.}~\bibnamefont {Fodor}},
  \bibinfo {author} {\bibfnamefont {S.~D.}\ \bibnamefont {Katz}}, \bibinfo
  {author} {\bibfnamefont {A.}~\bibnamefont {P\'asztor}}, \bibinfo {author}
  {\bibfnamefont {C.}~\bibnamefont {Ratti}},\ and\ \bibinfo {author}
  {\bibfnamefont {K.~K.}\ \bibnamefont {Szab\'o}},\ }\bibfield  {title}
  {\bibinfo {title} {Fluctuations and correlations in high temperature qcd},\
  }\href {https://doi.org/10.1103/PhysRevD.92.114505} {\bibfield  {journal}
  {\bibinfo  {journal} {Phys. Rev. D}\ }\textbf {\bibinfo {volume} {92}},\
  \bibinfo {pages} {114505} (\bibinfo {year} {2015})}\BibitemShut {NoStop}%
\bibitem [{\citenamefont {{Banerjee}}\ and\ \citenamefont
  {{Jedamzik}}(2004)}]{Banerjee2004}%
  \BibitemOpen
  \bibfield  {author} {\bibinfo {author} {\bibfnamefont {R.}~\bibnamefont
  {{Banerjee}}}\ and\ \bibinfo {author} {\bibfnamefont {K.}~\bibnamefont
  {{Jedamzik}}},\ }\bibfield  {title} {\bibinfo {title} {{Evolution of cosmic
  magnetic fields: From the very early Universe, to recombination, to the
  present}},\ }\href {https://doi.org/10.1103/PhysRevD.70.123003} {\bibfield
  {journal} {\bibinfo  {journal} {\prd}\ }\textbf {\bibinfo {volume} {70}},\
  \bibinfo {eid} {123003} (\bibinfo {year} {2004})},\ \Eprint
  {https://arxiv.org/abs/astro-ph/0410032} {arXiv:astro-ph/0410032 [astro-ph]}
  \BibitemShut {NoStop}%
\bibitem [{\citenamefont {{Trivedi}}\ \emph {et~al.}(2010)\citenamefont
  {{Trivedi}}, \citenamefont {{Subramanian}},\ and\ \citenamefont
  {{Seshadri}}}]{Trivedi2010}%
  \BibitemOpen
  \bibfield  {author} {\bibinfo {author} {\bibfnamefont {P.}~\bibnamefont
  {{Trivedi}}}, \bibinfo {author} {\bibfnamefont {K.}~\bibnamefont
  {{Subramanian}}},\ and\ \bibinfo {author} {\bibfnamefont {T.~R.}\
  \bibnamefont {{Seshadri}}},\ }\bibfield  {title} {\bibinfo {title}
  {{Primordial magnetic field limits from cosmic microwave background
  bispectrum of magnetic passive scalar modes}},\ }\href
  {https://doi.org/10.1103/PhysRevD.82.123006} {\bibfield  {journal} {\bibinfo
  {journal} {\prd}\ }\textbf {\bibinfo {volume} {82}},\ \bibinfo {eid} {123006}
  (\bibinfo {year} {2010})},\ \Eprint {https://arxiv.org/abs/1009.2724}
  {arXiv:1009.2724 [astro-ph.CO]} \BibitemShut {NoStop}%
\bibitem [{\citenamefont {Jedamzik}\ and\ \citenamefont
  {Saveliev}(2019)}]{Jedamzik2019}%
  \BibitemOpen
  \bibfield  {author} {\bibinfo {author} {\bibfnamefont {K.}~\bibnamefont
  {Jedamzik}}\ and\ \bibinfo {author} {\bibfnamefont {A.}~\bibnamefont
  {Saveliev}},\ }\bibfield  {title} {\bibinfo {title} {Stringent limit on
  primordial magnetic fields from the cosmic microwave background radiation},\
  }\href {https://doi.org/10.1103/PhysRevLett.123.021301} {\bibfield  {journal}
  {\bibinfo  {journal} {Phys. Rev. Lett.}\ }\textbf {\bibinfo {volume} {123}},\
  \bibinfo {pages} {021301} (\bibinfo {year} {2019})}\BibitemShut {NoStop}%
\bibitem [{\citenamefont {Jedamzik}\ and\ \citenamefont
  {Pogosian}(2020)}]{Jedamzik2020}%
  \BibitemOpen
  \bibfield  {author} {\bibinfo {author} {\bibfnamefont {K.}~\bibnamefont
  {Jedamzik}}\ and\ \bibinfo {author} {\bibfnamefont {L.}~\bibnamefont
  {Pogosian}},\ }\bibfield  {title} {\bibinfo {title} {Relieving the hubble
  tension with primordial magnetic fields},\ }\href
  {https://doi.org/10.1103/PhysRevLett.125.181302} {\bibfield  {journal}
  {\bibinfo  {journal} {Phys. Rev. Lett.}\ }\textbf {\bibinfo {volume} {125}},\
  \bibinfo {pages} {181302} (\bibinfo {year} {2020})}\BibitemShut {NoStop}%
\bibitem [{\citenamefont {{Wagstaff}}\ \emph {et~al.}(2014)\citenamefont
  {{Wagstaff}}, \citenamefont {{Banerjee}}, \citenamefont {{Schleicher}},\ and\
  \citenamefont {{Sigl}}}]{Wagstaff2014}%
  \BibitemOpen
  \bibfield  {author} {\bibinfo {author} {\bibfnamefont {J.~M.}\ \bibnamefont
  {{Wagstaff}}}, \bibinfo {author} {\bibfnamefont {R.}~\bibnamefont
  {{Banerjee}}}, \bibinfo {author} {\bibfnamefont {D.}~\bibnamefont
  {{Schleicher}}},\ and\ \bibinfo {author} {\bibfnamefont {G.}~\bibnamefont
  {{Sigl}}},\ }\bibfield  {title} {\bibinfo {title} {{Magnetic field
  amplification by the small-scale dynamo in the early Universe}},\ }\href
  {https://doi.org/10.1103/PhysRevD.89.103001} {\bibfield  {journal} {\bibinfo
  {journal} {\prd}\ }\textbf {\bibinfo {volume} {89}},\ \bibinfo {eid} {103001}
  (\bibinfo {year} {2014})},\ \Eprint {https://arxiv.org/abs/1304.4723}
  {arXiv:1304.4723 [astro-ph.CO]} \BibitemShut {NoStop}%
\bibitem [{\citenamefont {{Achikanath Chirakkara}}\ \emph
  {et~al.}(2021)\citenamefont {{Achikanath Chirakkara}}, \citenamefont
  {{Federrath}}, \citenamefont {{Trivedi}},\ and\ \citenamefont
  {{Banerjee}}}]{Chirakkara2021}%
  \BibitemOpen
  \bibfield  {author} {\bibinfo {author} {\bibfnamefont {R.}~\bibnamefont
  {{Achikanath Chirakkara}}}, \bibinfo {author} {\bibfnamefont
  {C.}~\bibnamefont {{Federrath}}}, \bibinfo {author} {\bibfnamefont
  {P.}~\bibnamefont {{Trivedi}}},\ and\ \bibinfo {author} {\bibfnamefont
  {R.}~\bibnamefont {{Banerjee}}},\ }\bibfield  {title} {\bibinfo {title}
  {{Efficient Highly Subsonic Turbulent Dynamo and Growth of Primordial
  Magnetic Fields}},\ }\href {https://doi.org/10.1103/PhysRevLett.126.091103}
  {\bibfield  {journal} {\bibinfo  {journal} {\prl}\ }\textbf {\bibinfo
  {volume} {126}},\ \bibinfo {eid} {091103} (\bibinfo {year} {2021})},\ \Eprint
  {https://arxiv.org/abs/2101.08256} {arXiv:2101.08256 [astro-ph.HE]}
  \BibitemShut {NoStop}%
\bibitem [{\citenamefont {Burrows}(1990)}]{Burrows1990}%
  \BibitemOpen
  \bibfield  {author} {\bibinfo {author} {\bibfnamefont {A.}~\bibnamefont
  {Burrows}},\ }\bibfield  {title} {\bibinfo {title} {Neutrinos from supernova
  explosions},\ }\href {https://doi.org/10.1146/annurev.ns.40.120190.001145}
  {\bibfield  {journal} {\bibinfo  {journal} {Annual Review of Nuclear and
  Particle Science}\ }\textbf {\bibinfo {volume} {40}},\ \bibinfo {pages} {181}
  (\bibinfo {year} {1990})},\ \Eprint
  {https://arxiv.org/abs/https://doi.org/10.1146/annurev.ns.40.120190.001145}
  {https://doi.org/10.1146/annurev.ns.40.120190.001145} \BibitemShut {NoStop}%
\bibitem [{\citenamefont {{Janka}}(2017)}]{Janka2017}%
  \BibitemOpen
  \bibfield  {author} {\bibinfo {author} {\bibfnamefont {H.-T.}\ \bibnamefont
  {{Janka}}},\ }\bibinfo {title} {{Neutrino Emission from Supernovae}},\ in\
  \href {https://doi.org/10.1007/978-3-319-21846-5\_4} {\emph {\bibinfo
  {booktitle} {Handbook of Supernovae}}},\ \bibinfo {editor} {edited by\
  \bibinfo {editor} {\bibfnamefont {A.~W.}\ \bibnamefont {{Alsabti}}}\ and\
  \bibinfo {editor} {\bibfnamefont {P.}~\bibnamefont {{Murdin}}}}\ (\bibinfo
  {year} {2017})\ p.\ \bibinfo {pages} {1575}\BibitemShut {NoStop}%
\bibitem [{\citenamefont {{Lesgourgues}}\ and\ \citenamefont
  {{Pastor}}(1999)}]{Lesgourgues1999}%
  \BibitemOpen
  \bibfield  {author} {\bibinfo {author} {\bibfnamefont {J.}~\bibnamefont
  {{Lesgourgues}}}\ and\ \bibinfo {author} {\bibfnamefont {S.}~\bibnamefont
  {{Pastor}}},\ }\bibfield  {title} {\bibinfo {title} {{Cosmological
  implications of a relic neutrino asymmetry}},\ }\href
  {https://doi.org/10.1103/PhysRevD.60.103521} {\bibfield  {journal} {\bibinfo
  {journal} {\prd}\ }\textbf {\bibinfo {volume} {60}},\ \bibinfo {eid} {103521}
  (\bibinfo {year} {1999})},\ \Eprint {https://arxiv.org/abs/hep-ph/9904411}
  {arXiv:hep-ph/9904411 [hep-ph]} \BibitemShut {NoStop}%
\bibitem [{\citenamefont {Dolgov}\ \emph {et~al.}(2002)\citenamefont {Dolgov},
  \citenamefont {Hansen}, \citenamefont {Pastor}, \citenamefont {Petcov},
  \citenamefont {Raffelt},\ and\ \citenamefont {Semikoz}}]{Dolgov2002b}%
  \BibitemOpen
  \bibfield  {author} {\bibinfo {author} {\bibfnamefont {A.}~\bibnamefont
  {Dolgov}}, \bibinfo {author} {\bibfnamefont {S.}~\bibnamefont {Hansen}},
  \bibinfo {author} {\bibfnamefont {S.}~\bibnamefont {Pastor}}, \bibinfo
  {author} {\bibfnamefont {S.}~\bibnamefont {Petcov}}, \bibinfo {author}
  {\bibfnamefont {G.}~\bibnamefont {Raffelt}},\ and\ \bibinfo {author}
  {\bibfnamefont {D.}~\bibnamefont {Semikoz}},\ }\bibfield  {title} {\bibinfo
  {title} {Cosmological bounds on neutrino degeneracy improved by flavor
  oscillations},\ }\href
  {https://doi.org/https://doi.org/10.1016/S0550-3213(02)00274-2} {\bibfield
  {journal} {\bibinfo  {journal} {Nuclear Physics B}\ }\textbf {\bibinfo
  {volume} {632}},\ \bibinfo {pages} {363} (\bibinfo {year}
  {2002})}\BibitemShut {NoStop}%
\end{thebibliography}%
	
\end{document}